\documentclass[
reprint,
superscriptaddress,
nofootinbib,
amsmath,amssymb,preprintnumbers,
aps,prd,
]{revtex4-2}

\usepackage{graphicx}% Include figure files
\usepackage{verbatim}
\usepackage{dcolumn}% Align table columns on decimal point
\usepackage{bm}% bold math
\DeclareUnicodeCharacter{2299}{\odot}
\usepackage{hyperref}% add hypertext capabilities
\usepackage{stackengine}
\usepackage{booktabs}
 \hypersetup{
     colorlinks= true,
     linkcolor = red,
     citecolor = red,      
     }
\usepackage{tikz-feynman,contour}

\begin{document}

\preprint{QMUL-PH-25-22}
\title{Resummed energy loss in extreme-mass-ratio scattering using critical orbits}

\author{Leor Barack}\email{L.Barack@soton.ac.uk} 
\affiliation{Mathematical Sciences, University of Southampton, Southampton, SO17 1BJ, United Kingdom}
\author{Riccardo Gonzo}\email{r.gonzo@qmul.ac.uk} 
\affiliation{Centre for Theoretical Physics, Department of Physics and Astronomy,
Queen Mary University of London, London E1 4NS, United Kingdom}
\author{Benjamin Leather}\email{B.J.Leather@soton.ac.uk} 
\affiliation{Mathematical Sciences, University of Southampton, Southampton, SO17 1BJ, United Kingdom}%
\author{Oliver Long}\email{oliver.long@aei.mpg.de} 
\affiliation{Max Planck Institute for Gravitational Physics (Albert Einstein Institute), D-14476 Potsdam, Germany}%
\author{Niels Warburton}\email{niels.warburton@ucd.ie} 
\affiliation{School of Mathematics \& Statistics, University College Dublin, Belfield, Dublin 4, Ireland}%

\begin{abstract}
Motivated by recent efforts to bridge between weak-field and strong-field descriptions of black-hole binary dynamics, we develop a resummation scheme for post-Minkowskian radiative observables in extreme-mass-ratio scattering, augmented with post-Newtonian terms. Specifically, we derive universal interpolation formulas for the total energy emitted in gravitational waves out to infinity and down the event horizon of the large black hole, valid to leading order in the small mass ratio.
We test our formulas using numerical results from direct calculations in black hole perturbation theory. The central idea of our approach is to utilize as a strong-field diagnostic the known form of divergence in the radiated energy along geodesics near the parameter-space separatrix between scattering and plunge. The dominant, logarithmic term of this divergence can be expressed in terms of instantaneous energy fluxes calculated along the unstable circular geodesics that form the separatrix, fluxes that we obtain using interpolation of highly accurate numerical data. The same idea could be applied to bound-orbit radiative observables via either unbound-to-bound mapping or a direct resummation of bound-orbit post-Newtonian expressions.   
\end{abstract}

\maketitle

%----------------------------------------------------------------------
\section{Introduction}\label{sec:intro}
%----------------------------------------------------------------------

Since its 2015 breakthrough, gravitational-wave astronomy has advanced at a breathtaking pace. The LIGO-Virgo-KAGRA (LVK) Collaboration has already cataloged more than three hundred compact binary signals \cite{LIGOScientific:2025slb}, turning once-rare transients into a population science. With the Laser Interferometer Space Antenna (LISA) on the horizon \cite{LISA:2024hlh}, the accessible spectrum will extend to massive black hole mergers and extreme mass-ratio inspirals (EMRIs), promising an unprecedented wave of discoveries in astrophysics, cosmology and fundamental physics. Precision modeling of gravitational wave sources forms an essential part of this program. With the rapid advance in detector technology, we are nearing a stage where the bottleneck in our ability to extract physics from observations is no longer detector sensitivity but rather systematic source modeling errors. The case of GW231123 \cite{LIGOScientific:2025rsn} provides a stark illustration of the challenge.

Waveform models for gravitational wave astronomy are based on Effective One Body (EOB), phenomenological, and reduced-order modeling techniques, informed by full simulations in Numerical Relativity as well as perturbative calculations. In this context there exist several key perturbative frameworks, each tailored to a particular physical regime. In particular, the gravitational self-force (SF) program~\cite{Barack:2018yvs,Pound:2021qin} has emerged as the primary framework for modeling EMRIs, while the post-Minkowskian (PM) and post-Newtonian (PN) expansions have recently seen rapid progress in the weak-field, scattering regime of the relativistic two-body problem~\cite{Blanchet:2013haa,Buonanno:2022pgc,DiVecchia:2023frv}.
There is a strong motivation to develop synergistic treatments that blend together the various perturbative approaches in the appropriate way in order to produce models that are uniformly accurate across parameter space. Such synergies are not only of conceptual importance but also of direct practical use in waveform modeling: for instance, information extracted from PM calculations of the scattering angle for spinless binaries~\cite{Damour:2016gwp,Bern:2019nnu,Damour:2020tta,DiVecchia:2021bdo,Bern:2021dqo,Bern:2021yeh,Dlapa:2021npj,Dlapa:2021vgp,Dlapa:2024cje,Driesse:2024feo,Driesse:2024xad,Dlapa:2025biy,Bern:2025wyd,Driesse:2026qiz} has already been successfully incorporated into precision EOB models for black hole binary dynamics~\cite{Damour:2017zjx,Damour:2019lcq,Antonelli:2019ytb,Khalil:2022ylj,Buonanno:2024vkx,Buonanno:2024byg,Long:2025nmj,Damour:2025uka}. Conversely, EOB models initially constructed to generate bound orbit waveforms have been extended to the unbound case and used to explore the scatter-capture separatrix~\cite{Long:2025nmj,Albanesi:2024xus}. Beyond such applications, recent work has further uncovered a striking connection between scattering observables and bound-orbit quantities~\cite{Kalin:2019rwq,Cho:2021arx,Saketh:2021sri,Gonzo:2023goe,Adamo:2024oxy,Khalaf:2025jpt}, opening a new avenue for a direct translation of scattering results into predictions for bound inspirals.

The availability of several complementary asymptotic approximations can be exploited to inform an efficient {\it resummation} of the perturbative expansions. In the case of SF and PM, one can use the strong-field behavior of certain invariant quantities calculated in a SF expansion to guide a resummation of the weak-field PM series, to the effect of producing a useful, uniformly valid approximant. This was illustrated recently in the example of the scattering angle, utilizing its known form of divergence near the parameter-space separatrix between scattering and plunging orbits~\cite{Damour:2022ybd,Rettegno:2023ghr,Long:2024ltn}.

Here we turn our attention to radiative fluxes in a scattering process, with the aim of using similar ideas to obtain PM-resummed formulas for the total energy lost via emission of gravitational waves, focusing on Schwarzschild black holes and working at leading order in the binary mass ratio $q:=m/M$. 
We use a terminology whereby `$n$PM order' describes the $\mathcal{O}(G^{n})$ term of the energy loss (where $G$ is the gravitational constant), and `1SF' refers to its $O(q^2)$ term. 
In this notation, the total energy radiated to infinity is now fully known through 4PM order~\cite{Herrmann:2021tct,Herrmann:2021lqe,Riva:2021vnj,DiVecchia:2022piu,Dlapa:2022lmu,Damgaard:2023ttc,Jakobsen:2023hig} and through 1SF at 5PM order~\cite{Driesse:2024feo}. 
Horizon-absorption contributions first arise at 7PM order and have been calculated at this order~\cite{Goldberger:2020wbx,Jones:2023ugm,Cipriani:2026myb}. In the context of black-hole perturbation theory, the behavior of orbits near the separatrix and the properties of gravitational radiation from such orbits were analyzed in a variety of studies over the years, including~\cite{Cutler:1994,Glampedakis:2002ya,Gundlach:2012aj,Berti:2010ce}. Details of the divergent behavior particularly relevant for our current work have been explored in Ref.~\cite{Barausse:2021xes}.

In this work, we construct novel semi-analytic models for the total amount of energy radiated during a scattering encounter. These models combine PM expressions, augmented with PN terms, with SF results to describe the total energy carried away in gravitational radiation to infinity and down the black hole horizon, $E^+_{\rm GW}$ and $E^-_{\rm GW}$ respectively. To assess the accuracy of these models, we compare their predictions against numerical calculations for scattering geodesics in black-hole perturbation theory~\cite{Warburton:2025ymy}. As a diagnostic of the 1SF strong-field dynamics we use the singular behavior of $E^+_{\rm GW}$ and $E^-_{\rm GW}$ for scattering geodesics approaching the separatrix in the parameter space. A key idea is that, near the separatrix, radiative losses are dominated by radiation from the near-circular whirl motion at periastron; in the separatrix limit itself the number of whirls diverges, and with it the radiative losses formally diverge too. This divergence has the form
$E^\pm_{\rm GW}\propto \log(\delta j)$, where $\delta j$ is the difference between the (a-dimensionalized) 
orbital angular momentum and its critical value at a given geodesic energy,
{\it and the proportionality coefficient is entirely determined from the energy flux calculated on the asymptotic unstable circular geodesic orbit}. Such circular-orbit fluxes can be readily computed with great numerical precision using standard tools in black-hole perturbation theory, and in this work we will obtain an accurate analytical fit for these fluxes (to infinity and down the horizon) as functions along the separatrix. These analytical fit formulas then determine the coefficient of the logarithmic divergence term that informs our resummation procedure.

We begin in Sec.~\ref{sec:0SF} by revisiting relevant aspects of geodesic motion in Schwarzschild spacetime, with a focus on the singular behavior near the separatrix between scattering and plunging orbits. We also connect the origin of this divergent behavior with the analytic structure of the radial action. In Sec.~\ref{Sec:critical} we consider the energy lost in gravitational radiation by a scattering geodesic source (neglecting radiation reaction on the orbit), analyzing its divergent behavior in the separatrix limit and obtaining an asymptotic formula for it in terms of the energy flux along the limiting unstable circular geodesic. This formula later forms a crucial ingredient of our resummation procedure. 
In Sec.~\ref{sec:PNPM} we review other such ingredients, namely the currently available PM and PN results for the energy loss in scattering scenarios, and also construct a mixed PM/PN model. These weak-field approximation models are then tested against numerical results in black hole perturbation theory.
Section \ref{sec:resum} introduces our resummation procedure, designed to improve the performance of the weak-field formulas in the strong-field regime. 
The formula is tested in various ways in Sec.~\ref{Sec:results}, again using numerical results as a benchmark. Section~\ref{sec:conclusion} provides a summary and final commentary.

%----------------------------------------------------------------------
\textit{Conventions---} Throughout the rest of this work we set $c=1=G$, and we use the mostly plus $(-+++)$ signature convention for the metric. 
%----------------------------------------------------------------------

%----------------------------------------------------------------------
\section{Scattering geodesics and the separatrix}
\label{sec:0SF}

We begin by reviewing essential features of scattering geodesics in Schwarzschild spacetime, with focus on the behavior near the separatrix. 

We consider a pointlike particle of mass $m$ scattered off a Schwarzschild black hole of mass $M\gg m$. In the test-particle limit, $q := m/M \to 0$, the particle moves along a timelike geodesic of the Schwarzschild background. We use Schwarzschild coordinates $x^\alpha=(t,r,\theta,\varphi)$ adapted to the black hole, denote the particle's geodesic trajectory by $x^\alpha_{p}(\tau)$ with $\tau$ being proper time along the geodesic, and without loss of generality take the trajectory to lie in the equatorial plane, i.e.~$\theta_p\equiv\pi/2$. The particle's conserved energy and angular momentum per $m$ are given, respectively, by $E := -u_t>1$ and $L := u_\varphi$, where $u_{\alpha}=g_{\alpha\beta}u^\beta$, with $u^\beta:=dx_{p}^\beta/d\tau$ being the particle's four-velocity and $g_{\alpha\beta}$ being the background Schwarzschild metric. For later convenience we also introduce the
dimensionless angular momentum parameter 
\begin{equation}
    j:=L/M
\end{equation}
used prevalently in PN literature. 
The geodesic equations of motion then admit the first-integral form
\begin{align}
\dot{t}_p &= \frac{E}{f(r_p)}, \quad
\dot \varphi_p = \frac{j M}{r_p^2}, \quad
\dot{r}_p = \pm \sqrt{E^2 - V(r_p; j)}\, ,
\label{eq:eoms_geo}
\end{align}
where the overdot denotes $\mathrm{d} / \mathrm{d} \tau$, $f(r) := 1 - 2M/r$, and the effective radial potential is given by
\begin{align}
V(r; j) = f(r)\left(1 + j^2\frac{M^2}{r^2}\right).
\end{align}

The pair ($E,j$) can be used to parametrize the family of timelike geodesics in both scattering and bound-orbit cases. In either case, it is useful to consider 
the real roots $\{r_2,r_1,r_0\}$ of the cubic equation $E^2 = V(r; j)$, given by
\begin{align}
r_2 &= \frac{6M}{1 + 2\zeta \cos (\xi/3)}, \quad
r_1 = \frac{6M}{1 + 2\zeta \cos\left[ (2 \pi + \xi)/3 \right]}, \nonumber \\
& \qquad \qquad  r_0 = \frac{6M}{1 + 2\zeta \cos\left[ (2 \pi - \xi)/3 \right]},
\label{eq:turning_points}
\end{align}
with 
\begin{align}
\zeta &:= \sqrt{1 - 12/j^2}, \nonumber \\
\xi &:=
\arccos \left( \frac{1 + (36 - 54 E^2) /j^2}{\zeta^3} \right)\, ,
\label{eq:zeta-xi_def}
\end{align}
and with $2M<r_2<r_0$ and $r_1<0$. For a scattering geodesic, the motion is confined to $r\geq r_0$, with $r_0$ being the periapsis distance.
The above roots define notions of eccentricity $e>1$ and semilatus rectum $p$ via
\begin{align}
r_0=\frac{M p}{1+e}\,, \quad r_1=\frac{M p}{1-e}\,, \quad r_2=\frac{2 M p}{p-4}\,,
\label{eq:ep_EL_conversion}
\end{align}
with the pair $(e,p)$ providing an alternative, geometrical parameterization of scattering geodesics.
Another useful parameterization is provided by the magnitude of the 3-velocity at infinity (in the black hole's frame),
\begin{equation}\label{v}
    v = \frac{\sqrt{E^2-1}}{E},
\end{equation}
together with the impact parameter
\begin{equation}\label{b}
    b:=\lim_{\tau\to-\infty} r_p(\tau)\sin\Big|\varphi_p(\tau)-\varphi_p(-\infty)\Big|=\frac{jM}{vE}.
\end{equation}

The boundary between plunging and scattering geodesics defines a curve $j_{\rm c}(E)$ in the parameter space, called the \emph{separatrix}. It is obtained from the conditions
\begin{align}
\dot{r}_p = 0=\ddot{r}_p \quad \Rightarrow \quad  E^2 - V(r; j)=0= \partial_r V(r; j)\,,
\label{eq:separatrix_def}
\end{align}
together with $\partial_r^2 V(r; j) < 0$, which give
\begin{align}
\hspace{-5pt}j_{\rm c}(E) &=\sqrt{\frac{27 E^4-36 E^2+8+E \left(9 E^2-8\right)^{3/2}}{2 (E^2 - 1)}}\, ,
\label{eq:Lcrit}
\end{align}
or, in terms of $v$,
\begin{align}
\hspace{-5pt}j_{\rm c}(v) &=\frac{1}{v} \sqrt{\frac{8 v^4+20v^2-1+\left(8v^2+1\right)^{3/2}}{2 (1 - v^2)}}\, .
\label{Lcofv}
\end{align}
Geodesics with $j>j_{c}(v)$ scatter, while those with $j<j_c(v)$ plunge. 
The critical value of the corresponding impact parameter, at a given $v$, is 
\begin{equation}\label{b_c}
    b_{\rm c}(v) = \frac{j_{\rm c}(v)M}{v E}\,,
\end{equation}
with $E=(1-v^2)^{-1/2}$. On the separatrix the roots $r_2$ and $r_0$ in Eq.~\eqref{eq:turning_points} coincide, signaling the formation of an unstable circular geodesic orbit at the ``whirl'' radius
\begin{align}
R&= \frac{M p_{\rm c}(e)}{1+e} \nonumber \\
&= \frac{1}{2}Mj_{\rm c}^2(v) \left[ 1- \sqrt{1-12/j_c^2(v)} \right]\,,
\label{eq:circular_radius}
\end{align}
where $p_{\rm c}(e)=6+2e$.
\begin{figure}[thb]
    \centering
    \includegraphics[width=\linewidth]{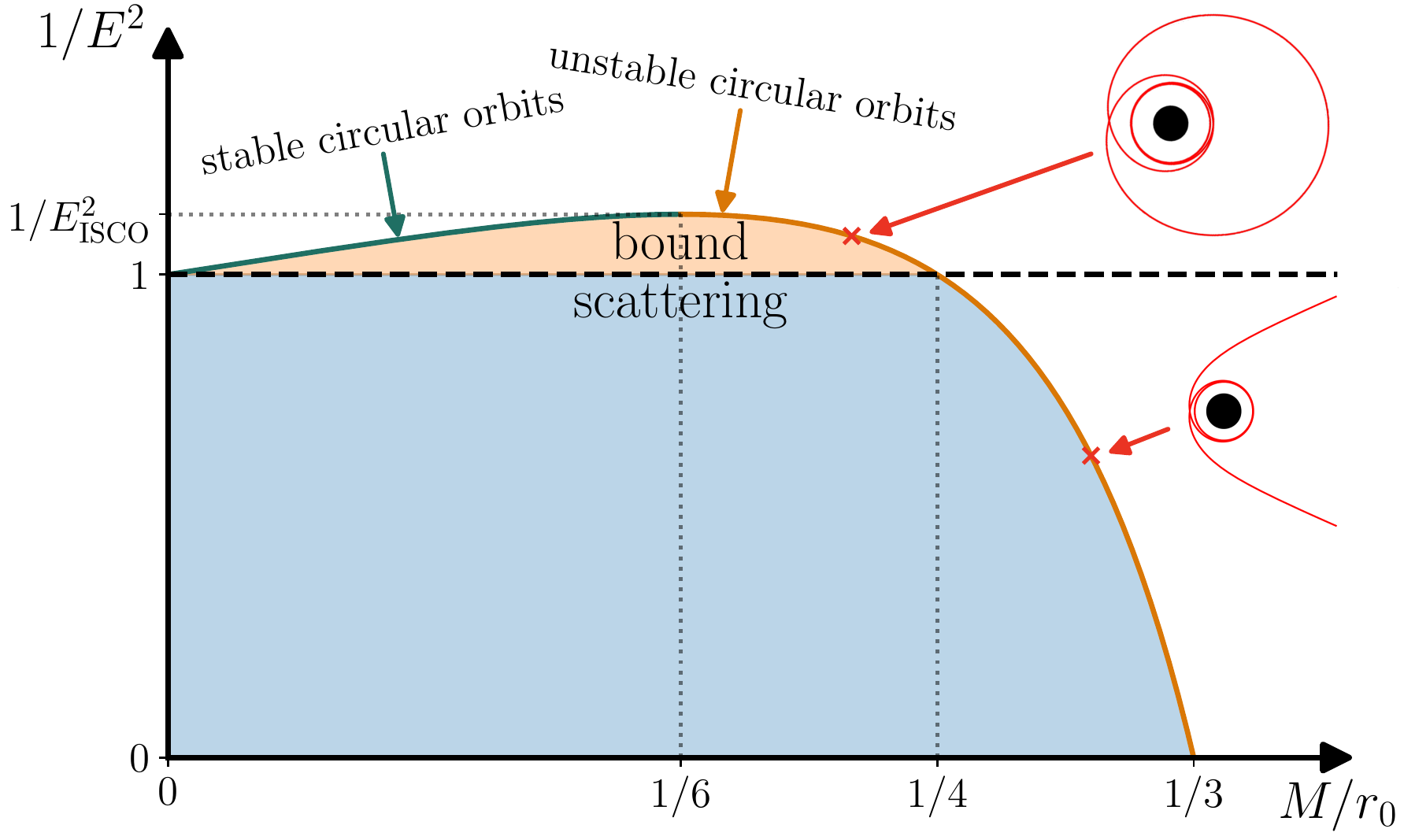}
    \caption{Representation of the space of (non-plunging) geodesic orbits around a Schwarzschild black hole, using the squared inverse specific orbital energy, $1/E^2$, and inverse pericenter radius, $M/r_0$, as compact coordinates on parameter space. Non-plunging geodesics consist of periodic/bound orbits ($1/E^2>1$) and scattering orbits ($1/E^2\leq 1$). The boundary of the region is formed by circular geodesics; it is made up of two segments, green and orange curves in the diagram, corresponding, respectively, to stable and unstable circular geodesics. The two segments join smoothly at the Innermost Stable Circular Orbit (ISCO). 
   The locus of unstable circular orbits (orange curve) is the separatrix, represented in the text by the function $r_0=R(E)$. Points along the separatrix also represent   homoclinic orbits (for $1/E^2>1$, $4M<R<6M$) or heteroclinic orbits (for $1/E^2<1$, $3M<R\leq 4M$), which display infinite ``zoom-whirl'' behavior. The two orbits illustrated are examples of near-separatrix homoclinic and heteroclinic geodesics [their parameters are $(E=0.962253, j=3.67426)$ and $(E=1.26486, j=5.77319)$, respectively]. Our resummation technique takes advantage of the singular behavior of geodesics near the separatrix as a marker of strong-field dynamics.  
    }
    \label{fig:separatrix}
\end{figure}
The monotonically decreasing function $R(v)$ provides a convenient parameterization of the \emph{separatrix curve}, depicted in Fig.~\ref{fig:separatrix}, with $3M<R<4M$ for scattering orbits.  The case $R=4M$ corresponds to the marginally-bound (``parabolic'') scattering geodesic with $v=0$~\cite{Barack:2019agd}, and $R\to\ 3M$ is the ultra-relativistic scattering limit $v\to 1$. 
In terms of the whirl radius $R$, the critical angular momentum in Eq.~\eqref{eq:Lcrit} and the corresponding circular-orbit specific energy take the simple forms 
\begin{equation}\label{LcofR}
    j_{\rm c}(R) = \frac{R/M}{\sqrt{R/M - 3}}\,, \quad E(R)=\frac{1-2M/R}{\sqrt{1-3M/R}}\,.
\end{equation}
It is interesting to note---cf.~Fig.~\ref{fig:separatrix}---how the separatrix extends smoothly to the bound-orbit regime $2\sqrt{2}/3\leq E<1$, where $4M<R\leq 6M$ and $R=6M$ is the innermost stable circular orbit (ISCO).

Near the separatrix, for a small $\delta j:= j-j_{\rm c}(v)>0$, scattering geodesics exhibit the so-called ``zoom-whirl'' behavior~\cite{Cutler:1994}, completing many azimuthal cycles in a near-circular motion just outside the periapsis distance. The total azimuthal phase traversed by the particle diverges logarithmically as $\delta j\to 0$, in accordance with
\begin{align}\label{log divergence}
    \Delta\varphi&:=\Big|\varphi_p(\tau\to\infty)-\varphi_p(\tau\to-\infty)\Big| \nonumber \\
    &\stackrel{\delta j \to 0}{\sim} A(v) \log\left(\frac{\delta j}{j_c(v)}\right) + {\rm const}(v)+\cdots 
\end{align}
where ${\rm const}(v)$ does not depend on $\delta j$ and the ellipsis denote terms that vanish in the separatrix limit. The coefficient of the logarithmic term in Eq.~\eqref{log divergence} is 
\begin{equation}\label{A}
A(v)=-\left(1-\frac{12}{j_c^2(v)}\right)^{-1/4},
\end{equation}
a derivation of which can be found in Appendix A of~\cite{Long:2024ltn}. In terms of the whirl radius $R$, this becomes
\begin{equation}\label{A(R)}
    A(R)=-\left(\frac{6M}{R}-1\right)^{-1/2}.
\end{equation}

\subsection{Separatrix singularity from a radial action}

It is instructive to consider the near-separatrix behavior also from the point of view of an action formulation, central to effective field theory and amplitude methods. 
The radial action for a scattering geodesic is given by 
\begin{align}
I^{(\bar{r})}_r(v, j) &= \frac{2}{M} \int_{r_{0}}^{\bar{r}} \frac{\sqrt{E^2 - V(r; j)}}{1-2M/r} dr\,, 
\label{eq:radial_action_def}
\end{align}
where the infrared regulator $\bar r$ is introduced for mathematical definiteness,  with $r_{0} \ll \bar r < \infty$, noting the integral formally diverges for $\bar r\to\infty$. The limit $\bar r \to \infty$ is taken only at the end of the calculation, after constructing a scattering ``observable'' from $I_r^{(\bar r)}$ (such as the scattering angle; see Eq.~\eqref{eq:chi_from_action} below), when one then aims to demonstrate that the result is finite.
For alternative ways of regulating the integral in Eq.~(\ref{eq:radial_action_def}), see, e.g., Ref.~\cite{Gonzo:2024xjk}.

For a small $\delta j>0$ at fixed $v$ , the radial action \eqref{eq:radial_action_def} admits an asymptotic expansion 
\begin{align}
\hspace{-6pt} I^{(\bar{r})}_r \!\stackrel{\delta j \to 0}{\sim}& c^{(\bar{r})}_{\rm log}(v) \delta j\,   \log \left(\frac{\delta j}{j_{c}(v)}  \right) \! + c^{(\bar{r})}_1(v) \delta j +\dots ,  
\end{align}
with some finite coefficients $c^{(\bar{r})}_{\rm log}(v)$ and $c^{(\bar{r})}_1(v)$. This vanishes in the limit $\delta j \to 0$, as expected of the radial action of a circular orbit. The presence of the logarithmic term traces back to a branch point in the complex $j$ plane, and the radial action develops a branch cut\footnote{Such information about the analytic structure of the scattering amplitude (closely related to $I^{(\bar{r})}_r$~\cite{Bern:2021dqo})---most notably, the emergence of a strong-field Regge cut in the $(v,j)$ plane which is non-analytic in the Newton's constant $G$---may also be useful in developing bootstrap approaches to classical gravitational scattering, where constraints from analyticity play a key role.} starting at the critical value \eqref{eq:Lcrit}. In the ultra-relativistic limit, $E \to \infty$, where explicit expressions for $I^{(\bar{r})}_r$  are available in terms of hypergeometric functions~\cite{Parnachev:2020zbr,Akpinar:2025huz}, it is easy to verify that such a branch cut starts at $j_{\rm c}(v) = 3 \sqrt{3} (1-v^2)^{-1/2}$.

The radial action \eqref{eq:radial_action_def} acts as a generating functional for the dynamics via the Hamilton–Jacobi formalism~\cite{Damour:1988mr}; see~\cite{Gonzo:2024xjk} for a recent discussion in the context of the first law of binary black hole mechanics.
 As an example, the geodesic scattering angle $\Delta\varphi$ can be obtained using 
 \begin{align}\label{eq:chi_from_action}
\Delta \varphi = \pi -\lim_{\bar{r} \to \infty} \frac{\partial I^{\bar{r}}_r(v, j)}{\partial j},
\end{align}
which is indeed finite, unlike the scattering radial action itself. Near the separatrix, in particular, the calculation recovers Eq.~\eqref{log divergence} with the correct coefficient $A(v)$. This universal divergent behavior characterizes other observables derived from the radial action, such as coordinate or proper-time delays \cite{Gonzo:2024xjk}.

\section{Energy loss in the critical limit} \label{Sec:critical}

\subsection{Singular form of the energy loss}

The radiative dynamics of binary black holes near the critical threshold have been studied in detail in a variety of contexts; see, for example, Refs.\ \cite{Pretorius:2007jn} (numerical relativity), \cite{Gundlach:2012aj} (critical phenomena in black hole perturbation theory), \cite{Akcay:2012ea} (black hole perturbation theory and EOB) and \cite{Barausse:2021xes} (the ultrarelativistic limit). For our purposes we need an analytical model of the total energy radiated in gravitational waves by a {\em fixed} scattering geodesic near criticality, i.e.~for $\delta j\ll 1$, ignoring radiation back-reaction on the geodesic. The assumption of geodesic motion with no back-reaction holds approximately (to leading order in the mass ratio) in the adiabatic regime where $\delta j$ is not too small, but must ultimately fail as $\delta j$ tends to zero,
because the radiated energy from the fixed geodesic grows unboundedly in the critical limit, as we discuss below. We will nonetheless keep the geodesic fixed in our analysis even as we take $\delta j\to 0$, which would fit our purpose here. In principle, this setup can be made mathematically consistent by replacing the retarded boundary conditions of the problem with ``time-symmetric'' ones, in which a commensurate amount of radiation is pumped in from past null infinity and out of the past horizon to exactly compensate for radiative losses. In our concluding remarks we will further comment on the validity of ignoring radiative corrections to the orbit in our scheme. 

Consider a near-critical scattering geodesic with parameters $(v,j)$ such that $0<\delta j=j-j_c(v)\ll 1$. The total energy lost via radiation as the particle of mass $m$ moves along that geodesic, $E_{\rm GW}= E^+_{\rm GW}+E^-_{\rm GW}$, is given by the balance-law formula \cite{Quinn:1999kj}
\begin{equation}
    E_{\rm GW}= -m\int_{-\infty}^{\infty} \dot{E}\, d\tau,
\end{equation}
where $\dot{E}=dE/d\tau\propto m$ is the momentary rate of geodesic energy dissipation, expressible in terms of the local gravitational self-force acting on the particle \cite{Barack:2018yvs}. We take $\tau=0$ to be the moment of periapsis passage, $r_p(0)=r_0$ with $\dot{r}_p(0)=0$. 
We split $E_{\rm GW}$ into a ``whirl'' contribution,
\begin{equation}
    E^{\rm whirl}_{\rm GW} := - m\int_{-\tau_1}^{\tau_1} \dot{E}\, d\tau ,
\end{equation}
and a ``zoom'' contribution, $E^{\rm zoom}_{\rm GW}:=E_{\rm GW}- E^{\rm whirl}_{\rm GW}$, where 
$\tau_1>0$ is such that $\delta r_p(\tau_1):=r_p(\tau_1)-r_0\ll M$. The geodesic equation of motion then implies $|\dot{r}_p(\tau)|\ll 1$ for $-\tau_1\leq \tau\leq \tau_1$, i.e.~ the motion remains nearly circular during the entire whirl. 

It is now useful to rewrite $E^{\rm whirl}_{\rm GW}$ in the form
\begin{equation}\label{E_whirl(phi)}
    E^{\rm whirl}_{\rm GW} =- m\int_{\varphi_p(-\tau_1)}^{\varphi_p(\tau_1)} (\dot{E}/\dot{\varphi}_p)\, d\varphi,
\end{equation}
where $\dot{E}$ and $\dot{\varphi}_p$ are considered functions of $\varphi$ along the scattering orbit. Both these functions approach constant, finite values in the limit $\delta j\to 0$, up to correction of $O(\delta r_p(\tau_1))$ that can be made arbitrarily small. At leading order in $\delta j$ and $\delta r_p(\tau_1)$, these values are the same constant values that these quantities admit on the limiting unstable circular geodesic of radius $R$. We denote these values by $\dot{E}_\circ(R)$ and
\begin{equation}\label{phi_circ}
\dot{\varphi}_{\circ}(R) =\frac{1}{R\sqrt{R/M-3}}.
\end{equation}
In this approximation, Eq. (\ref{E_whirl(phi)}) becomes
\begin{equation}\label{E_whirl}
    E^{\rm whirl}_{\rm GW} \simeq - m[\dot{E}_\circ(R)/\dot{\varphi}_{\circ}(R)] \Delta\varphi^{\rm whirl},
\end{equation}
where $\Delta\varphi^{\rm whirl}:=\varphi_p(\tau_1)-\varphi_p(-\tau_1)$ is the orbital phase accumulated during the whirl, which diverges in the limit $\delta j\to 0$. We see that  $E^{\rm whirl}_{\rm GW}$ inherits the singular behavior of $\Delta\varphi^{\rm whirl}$.

Finally, we note that the phase accumulated during the zoom, $\Delta\varphi-\Delta\varphi^{\rm whirl}$, is necessarily bounded for any finite $\tau_1$, and so is $E^{\rm zoom}_{\rm GW}$. This means that the singular relation (\ref{E_whirl(phi)}) applies, at leading order, also to the full quantities, i.e.~$E_{\rm GW} \simeq - m[\dot{E}_\circ(R)/\dot{\varphi}_{\circ}(R)] \Delta\varphi$. Substituting from Eqs.\ (\ref{phi_circ}), (\ref{log divergence}) and (\ref{A(R)}), we arrive at
\begin{align}\label{E_separatrix}
    E_{\rm GW} =& \frac{R^{3/2}\sqrt{R-3M}}{\sqrt{M(6M-R)}} m\dot E_{\circ}(R)\log\left(\frac{\delta j}{j_c(R)}\right)
    +{\rm const}(R)
    \nonumber\\ & +\cdots ,
\end{align}
where the ellipsis represents terms that vanish in the limit $\delta j\to 0$. Note the ($R$-dependent) constant term here has a nonlocal contribution coming from the entire zoom portion of the scattering orbit, so it cannot be obtained from a simple next-order local analysis near the unstable circular orbit. 

We can write down similar asymptotic formulas separately for the energy radiated to infinity and that absorbed by the black hole, $E_{\rm GW}^+$ and $E_{\rm GW}^-$ respectively. The well established energy balance law for bound orbits (see, e.g., \cite{Galtsov:1982hwm}), as applied to the limiting circular geodesic of radius $R$, implies 
\begin{equation}
-m\dot{E}_{\circ}(R)=(1-3M/R)^{-1/2}\left({\cal F}^+_{\circ}(R)+{\cal F}^-_{\circ}(R)\right),
\end{equation}
where the two terms on the right are respectively the fluxes of energy radiated in gravitational waves to null infinity and across the future event horizon, and the factor $(1-3M/R)^{-1/2}$ is $dt/d\tau$. Using this, Eq.~(\ref{E_separatrix}) splits into two corresponding contributions,
\begin{align}\label{E^pm_separatrix}
    E^\pm_{\rm GW} =& -\frac{R^{2}}{\sqrt{M(6M-R)}}\, {\cal F}^\pm_{\circ}(R)\log\left(\frac{\delta j}{j_c(R)}\right)+{\rm const}^{\pm}(R)
    \nonumber\\ &+\cdots .
\end{align}
The expressions in Eq.~(\ref{E^pm_separatrix}) (without the constant term) will be used in this study to model the divergent behavior of the radiated energies $E^\pm_{\rm GW}$ near the separatrix.

\subsection{Energy flux from unstable circular orbits: determination of ${\cal F}^\pm_{\circ}(R)$}

The circular-orbit fluxes ${\cal F}^\pm_{\circ}(R)$ in Eq.\ (\ref{E^pm_separatrix}) are not known in analytic form, but they can be calculated numerically with great precision for given values of $R$ using modern codes in black hole perturbation theory. Flux calculations are ubiquitous in the literature, but few have been performed for unstable circular orbits. Previous examples include Refs.~\cite{Gundlach:2012aj,Barausse:2021xes}, where numerical energy flux calculations were presented for circular geodesics below the ISCO.

Here we adopt the basic strategy of Ref.~\cite{Gundlach:2012aj} and compute the fluxes mode by mode in a Fourier-harmonic decomposition of the Lorenz-gauge metric perturbation equations sourced by a point mass on a circular geodesic with radius $3M<R\leq 6M$. We repeat the calculation for a dense sample of $R$ values in this range, in each case extracting the asymptotic fluxes at null infinity and across the future horizon.  
The calculation uses the code developed in \cite{Leather:2024mls}, which employs a hyperboloidal foliation together with spectral collocation and spatially compactified coordinates. This architecture enables the extraction of fluxes directly at null infinity and on the horizon, without incurring any extrapolation error. It also allows the computation to be efficiently parallelized across hundreds of Fourier modes, enabling us to achieve higher precision than in previous work. Such higher precision proved crucial, in particular, for capturing the logarithmic divergence of ${\cal F}^\pm_{\circ}$ near the light ring (see below). 

To obtain sufficient numerical precision in the total fluxes, it is necessary to compute and add together the contributions from a sufficient number of $\ell$-mode multipoles. The flux distribution into $\ell$ modes broadens significantly with decreasing $R$, compelling us to compute hundreds of mode contributions for $R$ values close to $3M$ in our sample. We employ an automated selection algorithm to determine an optimal truncation level $\ell_{\rm max}$, as follows. For each orbital radius $R$, and each $2\leq\ell\leq 500$, we compute the total flux per $\ell$-mode via 
${\cal F}^{\pm}_{\circ, \ell} = \sum^{\ell}_{m=-\ell} {\cal F}^{\pm}_{\circ, \ell m}$, where ${\cal F}^{\pm}_{\circ, \ell m}$ are individual $\ell,m$ contributions. At sufficiently large $\ell$, we expect an exponential fall-off, ${\cal F}^{\pm}_{\circ, \ell}\propto \exp(\kappa^\pm\ell)$, with some ($R$-dependent) $\kappa^\pm<0$. We apply a criterion to detect the onset of exponential behavior, and for each $\ell$ in the exponential tail, we extract the ``local'' exponent, $\kappa^\pm_\ell$, via a non-linear exponential fit,
and compute the error that would occur from neglecting the remaining piece of the exponential tail, using the formula for the sum of a geometric series:
\begin{equation}
\epsilon^\pm_{\rm trunc}={\cal F}^{\pm}_{\circ, \ell}\frac{e^{\kappa^\pm}}{1-e^{\kappa^\pm}}.
\end{equation}
Additional error in the partial sum over $\ell$-mode contributions comes from the numerical evaluation of the modal contributions ${\cal F}^{\pm}_{\circ, \ell}$ themselves. We crudely approximate it as
\begin{equation}
\epsilon^\pm_{\rm noise}=\epsilon\sum_{\ell=2}^{\ell}{\cal F}^{\pm}_{\circ, \ell}\, ,
\end{equation}
taking $\epsilon=10^{-12}$ as the effective machine precision in our calculation. The total error in the partial sum is taken to be $\epsilon^\pm_{\rm tot}=\max(\epsilon^\pm_{\rm trunc},\epsilon^\pm_{\rm noise})$, and the truncation mode $\ell_{\rm max}$ is then chosen to be the value of $\ell$ that minimizes $\epsilon_{\rm tot}$. This procedure is designed to ensure that we add all mode contributions at our disposal up to the point where the overall error in the partial sum over modes becomes dominated by numerical noise.  We use the partial sum $\sum^{\ell_{\rm max}}_{\ell=2} {\cal F}^{\pm}_{\circ, \ell}$ as our approximant for ${\cal F}^{\pm}_{\circ}$, and we record $\pm \epsilon_{\rm tot}$ as an error bar for it. Our circular-orbit flux dataset for $3M<R\leq 6M$ is available in the accompanying ancillary file \cite{ZenodoCirc}.

\begin{figure}[thb]
    \centering
    \includegraphics[width=\linewidth]{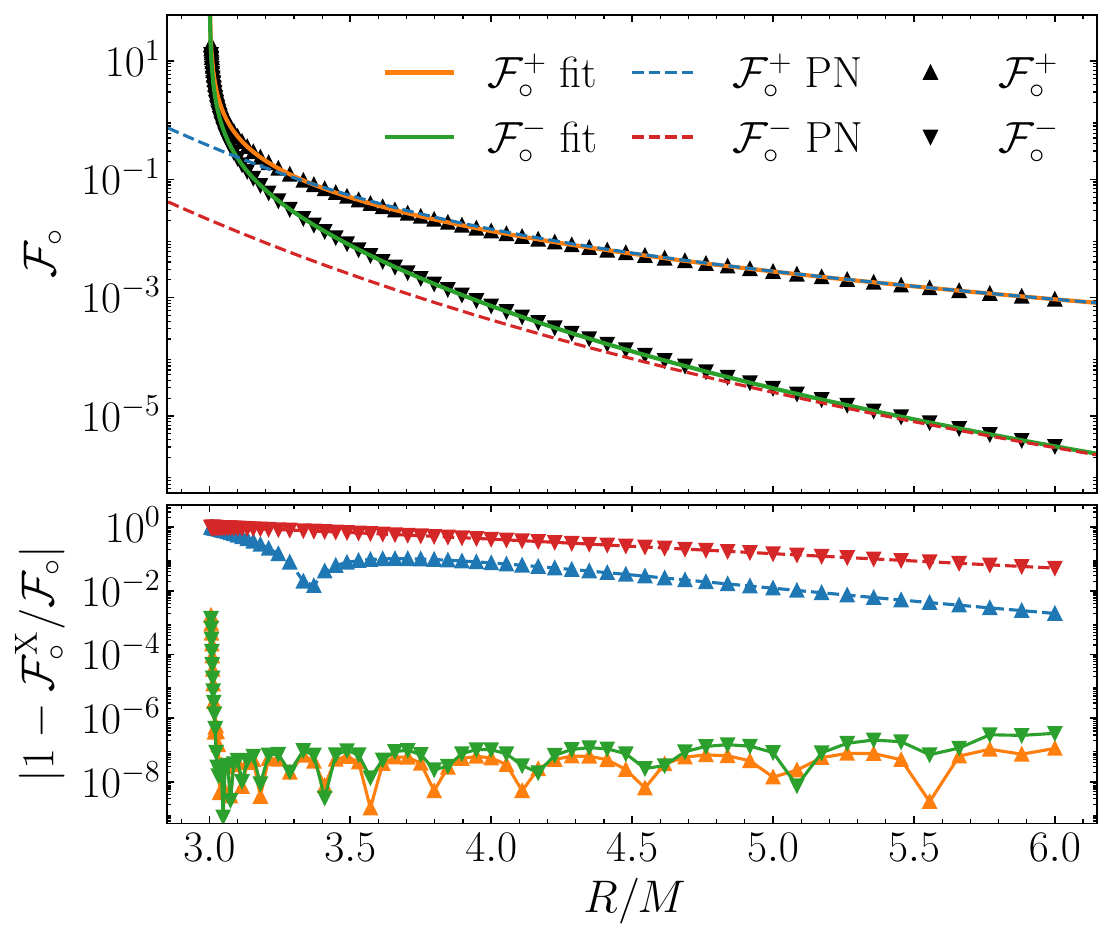}
    \caption{Energy flux (per $q^2$) from unstable circular geodesics, at null infinity (${\cal F}^+_{\circ}$) and down the event horizon (${\cal F}^-_{\circ}$), as a function of orbital radius $R$. Triangular markers show results from our high-precision numerical calculations (with error bars indiscernible on the scale of this plot).  Our analytical models (\ref{eq:circular_flux_fit}), with the best-fit parameters from Table~\ref{tab:flux_fit_coeffs}, are shown in solid lines. The corresponding PN expressions are shown in dashed line for comparison. The bottom panel displays the relative difference between each flux model and the numerical data.}
    \label{fig:F_circ}
\end{figure}

Next we fit our numerical ${\cal F}^{\pm}_{\circ}$ data to an analytical model. Following Ref.~\cite{Gundlach:2012aj} we introduce $z:=1-3M/R$, and note the expected divergent behavior ${\cal F}^{\pm}_{\circ}\sim (\log z)/z$ as $z\to 0$ ($R\to 3M$), predicted from the arguments and numerical evidence presented in Ref.\ \cite{Barausse:2021xes}. We also note the PN behavior 
${\cal F}^{+}_{\circ} \sim (1-z)^{5}$ 
and ${\cal F}^{-}_{\circ} \sim  (1-z)^{9}$ 
for $z\to 1$ [$R\to\infty$, assuming ${\cal F}^{+}_{\circ}(R)$ extends analytically to the stable circular-orbit regime, $R>6$]. With all this in mind, we fit our data to a rational-function model augmented with logarithmic terms, of the form 
\begin{equation}
    {\cal F}^{\pm}_{\circ}(z) = \frac{\sum^{5}_{i=0} z^{i}(a^{\pm}_{i} + b^{\pm}_{i}\log z)}{(1-z)^{-\alpha^{\pm}}\, z\, \big(1+\sum^{3}_{i=1} c^{\pm}_{i} z^{i}\big)}
    \label{eq:circular_flux_fit}
\end{equation}
with $\alpha^+=5$ and $\alpha^-=9$.
The coefficients $a^{\pm}_{i}$, $ b^{\pm}_{i}$ and $ c^{\pm}_{i}$ are determined via a non-linear least-squares optimization, with the polynomial orders selected to minimise residuals across the entire domain $0<z\leq 1/2$.  The resulting fit coefficients are listed in Table~\ref{tab:flux_fit_coeffs}. Figure \ref{fig:F_circ} displays the fitted functions ${\cal F}^\pm_{\circ}(R)$ with a comparison to the numerical data. Our models reproduce the data for ${\cal F}^\pm_{\circ}(R)$ to within a fractional difference of $< 10^{-6}$ over the range $3.02M \lesssim R\leq 6M$. (The domain $3M<R<3.02M$ corresponds to ultrarelativistic scattering with $v>0.97$, not explored in this work. Our fit could be refined near $R=3$ for the purpose of modeling the ultrarelativistic regime.)
\begin{table}[h]
    \centering
    \caption{Best-fit coefficients for the circular-orbit energy flux models in Eq.~(\ref{eq:circular_flux_fit}). }
    \label{tab:flux_fit_coeffs}
    \setlength{\tabcolsep}{6pt} % Slightly more padding for readability
    \begin{tabular}{c | r | r}
        \toprule
        Coeff. & \multicolumn{1}{c|}{${\cal F}^{-}_{\circ}(R)$} & \multicolumn{1}{c}{${\cal F}^{+}_{\circ}(R)$} \\
        \midrule
        $a_0$ & $-1.90224575 \times 10^{-2}$ & $-7.88200542 \times 10^{-3}$ \\
        $a_1$ & $-5.74081323 \times 10^{-1}$ & $-2.00088826 \times 10^{-2}$ \\
        $a_2$ & $-5.40484979 \times 10^{0}$ & $1.25312009\times 10^{0}$ \\
        $a_3$ & $-3.84335510 \times 10^{0}$ & $2.36546304 \times 10^{1}$ \\
        $a_4$ & $4.13644264 \times 10^{1}$ & $6.39305449 \times 10^{1}$ \\
        $a_5$ & $-3.12214636 \times 10^{1}$ & $-7.63501457 \times 10^{1}$ \\
        \midrule
        $b_0$ & $-6.96758492 \times 10^{-3}$ & $-7.09517088 \times 10^{-3}$ \\
        $b_1$ & $-2.20674209 \times 10^{-1}$ & $-9.31184640 \times 10^{-2}$ \\
        $b_2$ & $-2.59772993\times 10^{0} $ & $-1.72847811 \times 10^{-2}$ \\
        $b_3$ & $-1.20934561 \times 10^{1}$ & $5.57214378\times 10^{0}$ \\
        $b_4$ & $1.49920205 \times 10^{1}$ & $5.65614450 \times 10^{1}$ \\
        $b_5$ & $1.46307617 \times 10^{1}$ & $4.28876569 \times 10^{1}$ \\
        \midrule
        $c_1$ & $1.78346630 \times 10^{1}$ & $1.01401486 \times 10^{1}$ \\
        $c_2$ & $2.08383139\times 10^{0} $ & $1.78615448 \times 10^{1}$ \\
        $c_3$ & $7.80432921 \times 10^{2}$ & $4.70548933 \times 10^{2}$ \\
        \bottomrule
    \end{tabular}
\end{table}

As an extra check, we have compared our flux data to available analytical PN formulas; this too is shown in Fig.~\ref{fig:F_circ}. At $O(q^2)$ expressions have been derived to very high PN order through analysis of solutions to the Teukolsky equation. Here we used the 22PN-order expressions for ${\cal F}^+_{\circ}$ derived in~\cite{Tanaka:1996lfd,Fujita:2012cm}, and the 22.5PN-order expressions for ${\cal F}^-_{\circ}$ derived in~\cite{Poisson:1994yf,Fujita:2014eta}.  (Ref.~\cite{Cipriani:2025ikx} recently extended the calculation of ${\cal F}^+_{\circ}$ to 30PN.) These PN expressions also provide some insight into the analytical structure of ${\cal F}^\pm_{\circ}$ and support the functional form used in our fitting model, Eq.~(\ref{eq:circular_flux_fit}). It is striking to observe in Fig.~\ref{fig:F_circ} how well the PN formulas agree with the numerical data in the part of the domain nearer the ISCO, despite their being applied here deep in the strong-field regime.

\section{Energy loss in weak-field approximations} \label{sec:PNPM}

Our resummation method for $E_{\rm GW}^\pm$, to be described in the next section, will interpolate between the strong-field singular behavior described in Eq.~(\ref{E^pm_separatrix}) and analytical approximations based upon weak-field expansions. In this section we review the necessary weak-field input for this procedure, which involves results from calculations in both PM and PN theories.

\subsection{Energy loss in a Post-Minkowskian expansion}

In the weak-field regime,  $r_0\gg M$ (where, recall, $r_0$ is the periapsis distance), it is natural to apply a
PM expansion, i.e.~an expansion in  powers of $GM/r_0$ around a flat Minkowski background.  The total energy loss at infinity can be computed using the ``in-in'' formalism \cite{Schwinger:1960qe,Keldysh:1964ud}, 
either in a worldline effective theory \cite{Goldberger:2004jt} or in the amplitude-based KMOC approach \cite{Kosower:2018adc}, with Poincar\'e invariance fixing the form of the linearized stress–tensor operator that enters the quantization of the underlying field theory.  For the energy absorbed by the horizon, additional worldline degrees of freedom are required to encode the interaction between the horizon and the bulk fields \cite{Goldberger:2005cd}.  Their response is determined by matching to the low-frequency absorption properties of the black hole, allowing $E_{\rm GW}^-$ to be treated on the same footing as $E_{\rm GW}^+$ within the PM expansion.

At leading order in the mass ratio, we can write the PM expansion as
\begin{align}\label{E_PM}
    E_{\rm GW}^{\pm} &= M \nu^{2} \sum_{k = k_0^{\pm}}^\infty j^{-k}
    \mathcal{E}^{\pm}_{(k)}(v)\,,
\end{align}
where $\nu := m M/(m+M)^2= q +O(q^2)$ is the symmetric mass ratio (a more natural parameter used in lieu of $q$ in PM analysis with arbitrary mass ratios), and the leading expansion terms have $k_0^+=3$ and $k_0^-=7$. Recall $j$ is the conserved orbital angular momentum of the geodesic orbit (per $m$ and per $M$). The dimensionless coefficient functions $\mathcal{E}^{\pm}_{(k)}(v)$ encode the velocity dependence of the radiated and absorbed energy at $k$PM order. In this work, we use the current state-of-the-art PM results, which at $O(\nu^2)$ are 5PM for $E_{\rm GW}^+$ \cite{Driesse:2024feo} and 7PM (leading term) for $E_{\rm GW}^-$ \cite{Goldberger:2020wbx}. At leading PM order, the coefficients are given by
\begin{align}
\mathcal{E}^{+}_{(3)} &=
\frac{\pi}{48}
\Bigg[
\frac{6 E }{\sqrt{E^2-1}}\,
P_1(E)\, \sinh^{-1}\!\!\left(\sqrt{\frac{E-1}{2}}\right)
\nonumber\\
&\hspace{2em}
-\Big( P_2(E) +6\,P_3(E)\, \log\!\left(\frac{E+1}{2}\right) \Big) \Bigg] , \\
\mathcal{E}^{-}_{(7)} &= \frac{5 \pi}{16}
\left(21 E^4-14 E^2+1\right) (E^2-1)^4 , \label{E7}
\end{align}
with $E = (1-v^2)^{-1/2}$, and where  
\begin{align}
P_1(E) =& 70 E^6-165 E^4+112 E^2-33 ,  \nonumber\\
P_2(E) =& -210 E^6+552 E^5-339 E^4 +912 E^3 \nonumber \\
&-3148 E^2+3336 E-1151 , \nonumber \\
P_3(E) =& 35 E^6+60 E^5-185 E^4 +16 E^3+145 E^2
\nonumber\\
& -76 E+5 . \nonumber
\end{align}
The expressions for $\mathcal{E}^{+}_{(k>3)}$ are more unwieldy, and we refer the reader to \cite{Driesse:2024feo,Dlapa:2022lmu} for their explicit form. An implementation of these expressions is also available as the accompanying Mathematica script \cite{ZenodoMma}.

\subsection{Energy loss in a Post-Newtonian expansion}

A complementary weak-field approximation is obtained by expanding $E_{\rm GW}^\pm$ in powers of $v$, which is {\em a priori} useful in the small-$v$ regime of the scattering problem. This is similar to the familiar PN expansion for bound orbits, but unlike in the case of (say, circular) bound motion, here the small-$v$ limit does not automatically correspond to the weak-field limit $j\gg 1$, since $v$ and $j$ are independent parameters. In order to impose $j\to\infty$ for $v\to 0$, one can carry out the expansion in powers of $v$ while holding fixed the product $vj$, rather than $j$ itself.  
The ``PN limit'' of this expansion is then understood as one in which
\begin{equation}
v \to 0,
\quad
j \to \infty,
\quad
v j = \text{fixed},
\label{eq:scaling}
\end{equation}
ensuring an overlap with the PM domain of validity. 
In this scheme, the PN expansion of the $O(\nu^2)$ energy emitted to infinity takes the form
\begin{equation}\label{PN}
    \frac{E^+_{\rm GW}}{M\nu^2}=v^7\sum_{n=0}^\infty v^n \Big[\tilde a_n(vj)+ 
\!\!\sum_{i=1}^{\lfloor n/6\rfloor}\!\! \tilde b_{ni}(vj)\,(\log v)^i \Big] ,
\end{equation}
where logarithmic terms arise from hereditary tail effects, first manifest at (relative) 3PN order, i.e.~$n=6$. \cite{Blanchet:1992br,Blanchet:2013haa}

In reality, available PN results are not formulated directly in terms of an expansion at fixed $v j$, but instead they employ a quasi-Keplerian parameterization, in which the orbital dynamics are described in terms of a set of PN eccentricities. For hyperbolic orbits, one choice is the time eccentricity $e_t$ \cite{Blanchet:2013haa}, and usually PN calculations utilize the {\it proper} velocity at infinity, $p_\infty := E v$, as an expansion parameter instead of the coordinate velocity $v$. The leading, ``Newtonian-order'' form of the time eccentricity is 
\begin{equation}
e_t^2 \sim  e_N^2:=1 + (p_\infty j)^2, 
\label{eq:time-ecc}
\end{equation}
so the PN expansion at fixed $e_N$ is equivalent to one at fixed $p_\infty j=vj+O(v^3)$.
In particular, it too is automatically a weak-field expansion, so it is expected to have a domain of overlap with the PM expansion. 

In the above scheme we thus have the form 
\begin{equation}
\hspace{-8pt}\frac{E^+_{\rm GW}}{M \nu^2}
=
 p_\infty^7
\sum_{n=0}^{\infty}
\Big[
a_n(e_N)
+
\!\!\sum_{i=1}^{\lfloor n/6\rfloor}\!\!
b_{n,i}(e_N)\,(\log p_\infty)^i
\Big]
p_\infty^n .
\label{E_PN}
\end{equation}
The coefficient of the leading (0PN) term was derived already in the 1970s \cite{Hansen:1972jt,Turner:1977}; it is given by
\begin{align}
a_0(e_N)= 
& \frac{2}{45}
\frac{(602+673e_N^2)}{(e_N^2-1)^{3}}
\nonumber\\
&+\frac{2}{15}\cos^{-1}(1/e_N)\frac{(96+292e_N^2+37e_N^4)}{(e_N^2-1)^{7/2}}.
\label{eq:0PN}
\end{align}
Higher-order instantaneous coefficients up to 2PN order beyond the leading term are available in closed analytic form at arbitrary $e_N$ \cite{Blanchet:1989cu,Junker:1992kle,Bini:2021gat,Cho:2021onr}. At 2.5PN and 3PN order beyond the leading term ($n=5$ and $n=6$ respectively), the energy flux receives hereditary contributions involving nonlocal-in-time integrals. For hyperbolic motion, these integrals cannot be evaluated  in closed form at arbitrary eccentricity. However, they have been evaluated analytically term by term in an expansion in powers of $1/j$, i.e.~through a mixed PN-PM expansion.  The best currently available results at 2.5PN and 3PN orders include PM terms up to $O(j^{-15})$ \cite{Cho:2022pqy,Bini:2022enm}. We henceforth refer to this best available model as the $\rm 3'PN$ model,  where the prime serves to remind that $o(j^{-15})$ terms are omitted at 2.5PN and 3PN orders. The explicit $\rm 3'PN$ expressions used in the current work are provided in the accompanying Mathematica script \cite{ZenodoMma}.

Similar PN expansions are yet to be derived for the horizon-absorbed energy, so in this work we will be using only PM expressions for $E^-_{\rm GW}$.

\subsection{Energy loss in a mixed PM-PN expansion}

Combining the above PM and PN expansions in a suitable way, we can construct a weak-field approximation that is valid through both 5PM and 3PN orders, modulo $o(j^{-15})$ terms at 2.5PN and 3PN orders.  Let $E^+_{\rm 5PM}$ and $E^+_{\rm 3'PN}$ denote our 5PM and $\rm 3'PN$ approximations for $E^+_{\rm GW}$ [Eqs.\ (\ref{E_PM}) and (\ref{E_PN}) respectively]. We construct a mixed approximation by adding $E^+_{\rm 5PM}$ and $E^+_{\rm 3'PN}$ together and removing all overlapping terms:
\begin{equation}\label{mixed}
E^+_{\rm 5PM/3'PN} = E^+_{\rm 5PM} + E^+_{\rm 3'PN} - \left(E^+_{\rm 3'PN}\right)_{\rm 5PM}.
\end{equation}
Here $(E^+_{\rm 3'PN})_{\rm 5PM}$ is the result of PM-expanding $E^+_{\rm 3'PN}$ in powers of $1/j$ at fixed $p_{\infty}$ up to 5PM order [$O(j^{-5})$], i.e.,
\begin{align}
\label{eq:overlap_taylor}
&\left(E^+_{\rm 3'PN}\right)_{\rm 5PM}= \\
&\quad
\sum_{k=3}^{5}
\frac{1}{k! j^k}
\left[
\frac{\partial^k}{\partial x^k}  E^+_{\rm 3'PN}\left(p_{\infty},e_N(p_{\infty},j=1/x)\right)\right]_{x=0}\,. \nonumber 
\end{align}
This must be identical to $(E^+_{\rm 5PM})_{\rm 3PN}$---the result of PN-expanding $E^+_{\rm 5PM}$ up to 3PN order---and it captures exactly the overlapping terms that need to be subtracted to avoid double-counting. By construction, $E^+_{\rm 5PM/3'PN}$ is valid through both 5PM and $\rm 3'PN$ orders.

The explicit expression for $E^+_{\rm 5PM/3'PN}$ is provided in the accompanying Mathematica script \cite{ZenodoMma}.

\subsection{Comparison with numerical data}

\begin{figure}[tb]
    \centering
    \includegraphics[width=\linewidth]{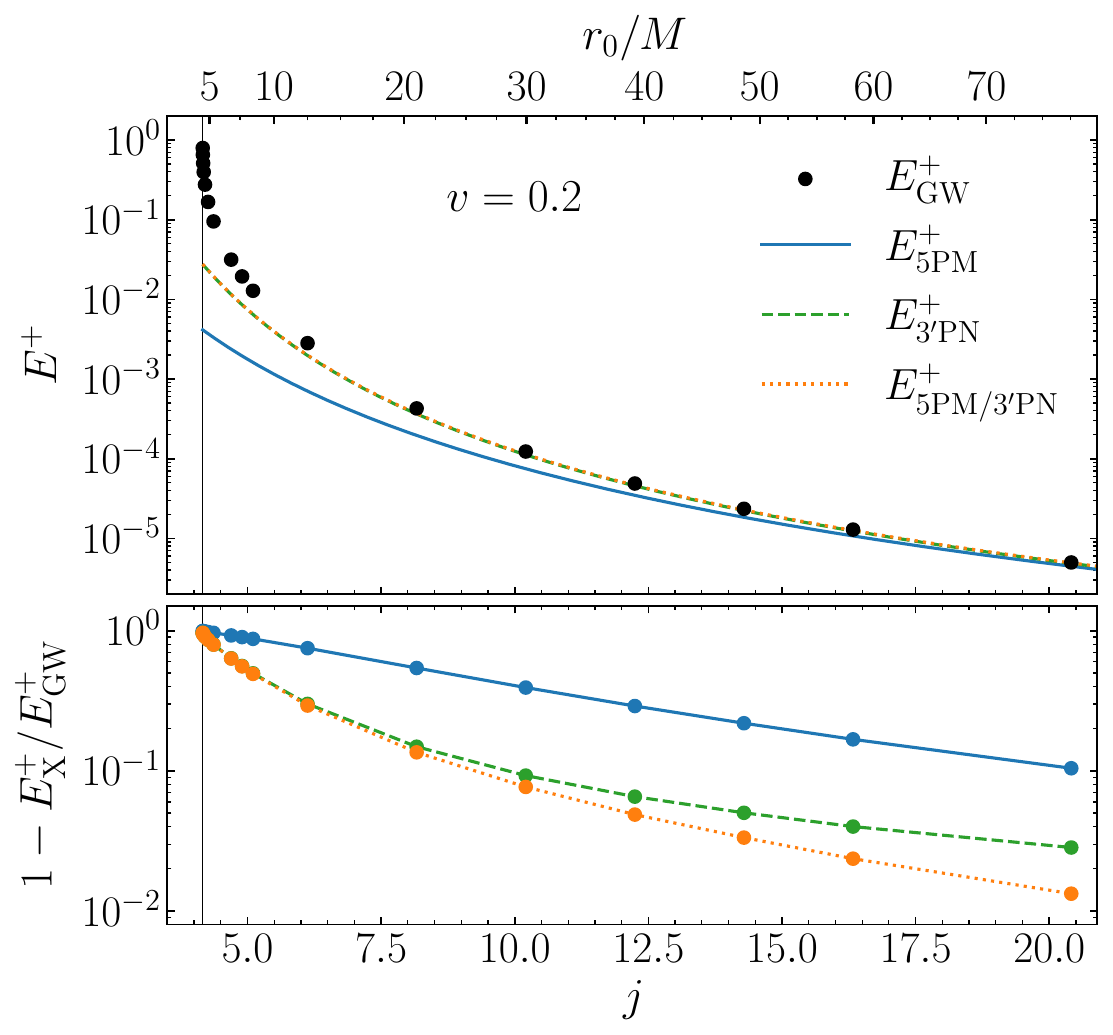}
    \caption{Total energy radiated to infinity in the weak-field approximations 
$E^+_{\rm 5PM}$, $E^+_{\rm 3'PN}$ and $E^+_{\rm 5PM/3'PN}$, as a function of orbital angular momentum $j$ (lower scale) and periapsis distance $r_0$ (upper scale) at a fixed value $v=0.2$ of the velocity at infinity. ``Exact'' benchmark results $E^+_{\rm GW}$ from the numerical code of Ref.\ \cite{Warburton:2025ymy} are shown for reference. (Estimated error bars on the numerical values are too small to be visible on the scale of this plot.) The lower panel displays relative differences with respect to the numerical data. The location of the separatrix is marked by the vertical line. Here and in all subsequent plots, the displayed quantities $E^{\pm}$ are normalized by a factor $q^2M(=m^2/M)$.}
    \label{fig:weak_field_0.2}
\end{figure}

\begin{figure}[tb]
    \centering
    \includegraphics[width=\linewidth]{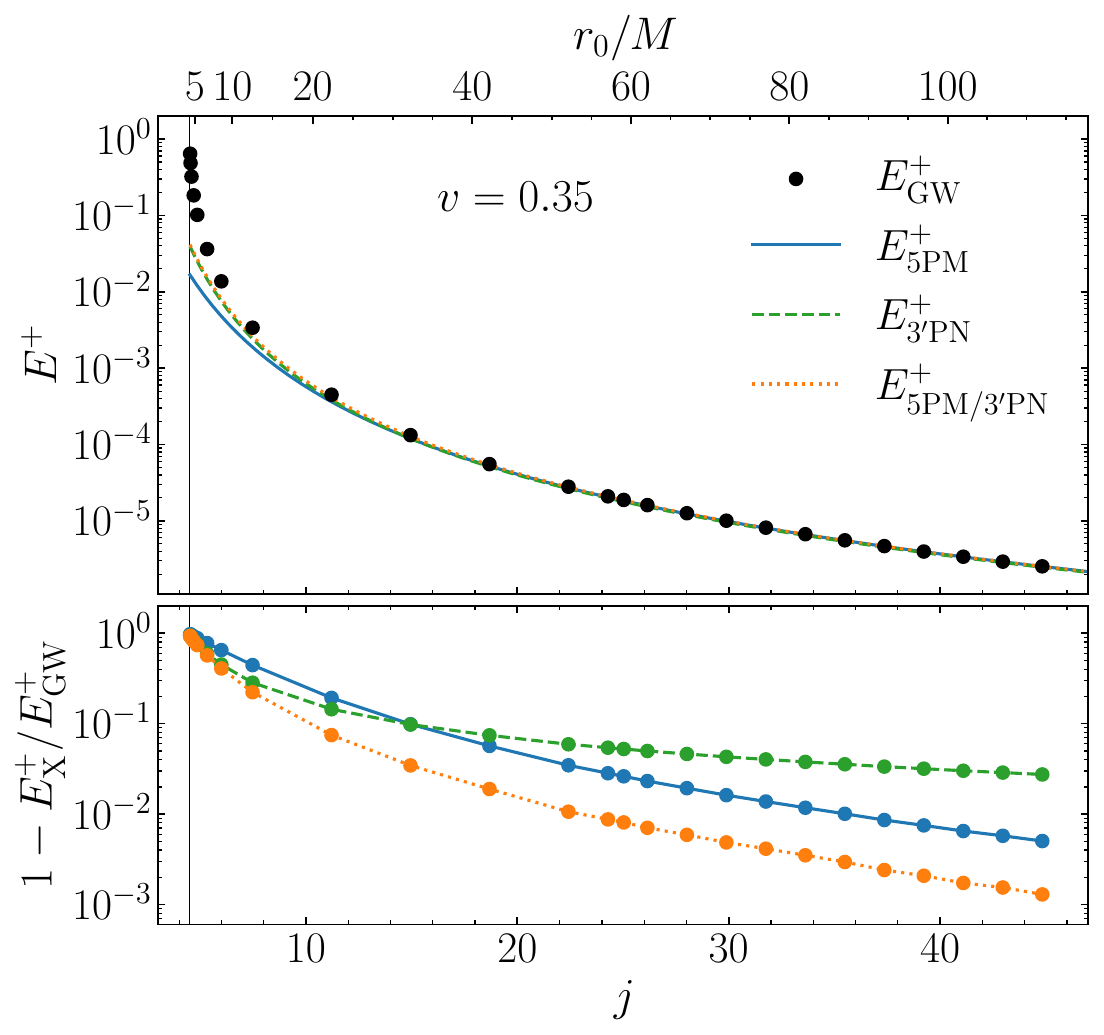}
    \caption{Same as Figure \ref{fig:weak_field_0.2}, this time for $v=0.35$. We see that the 5PM formula takes over from the $\rm 3'PN$ model as a better approximation at large $j$. The mixed 5PM/$\rm 3'PN$ model still outperforms both pure PM and pure PN models at all $j$. }
    \label{fig:weak_field_0.35}
\end{figure}

We can check the fidelity of each of our three weak-field approximations, $E^+_{\rm 5PM}$, $E^+_{\rm 3'PN}$ and $E^+_{\rm 5PM/3'PN}$,
by comparison with numerical results for $E^+_{\rm GW}$ obtained in black-hole perturbation theory. These benchmark results were recently computed by one of the authors (NW) in Ref.~\cite{Warburton:2025ymy}
using numerical integration of the Regge-Wheeler perturbation equations with a scattering geodesic source in the frequency-domain. Figures \ref{fig:weak_field_0.2} and \ref{fig:weak_field_0.35} illustrate the behavior of our approximations against benchmark data from the code of \cite{Warburton:2025ymy}, which are available in the accompanying ancillary file \cite{ZenodoRad}. We see that all three approximations work well in the weak-field regime, as expected. It is also demonstrated that the $\rm 3'PN$ formula outperforms the 5PM model at low velocity, but the situation reverses at higher $v$. Most notably, the mixed PM/PN model outperforms either of the two separate expansions across the entire domain and for both values of velocity shown. Finally, also as expected, we see how all three approximations ultimately break down in the strong-field regime. It is this breakdown of the weak-field approximations that we seek to cure with our resummation technique, to be introduced next.

\section{Resummation procedure}\label{sec:resum}

We seek analytical models for $E_{\rm GW}^{\pm}$ that smoothly interpolate between the weak-field approximations of the previous section and the strong-field singular behavior at the separatrix described in Section \ref{Sec:critical}.  More specifically, we look for a ``resummation'' of the weak-field expansions that reproduces the known logarithmic divergence at $j=j_c(v)$ without impacting the form of any known PM or PN expansion terms. We achieve this using the technique already mentioned: We first add the logarithmic divergence term (\ref{E^pm_separatrix}) to the appropriate weak-field expansion, and then remove common expansion terms. For instance, to resum the 5PM series (\ref{E_PM}) for $E^+_{\rm GW}$, we add it to the logarithmic term in (\ref{E^pm_separatrix}) and then subtract all terms in the PM expansion of the logarithmic term up to and including 5PM order. Since the subtracted ``counter-terms'' are regular at the separatrix, the resulting resummed expression is guaranteed to possess both the correct singular behavior at the separatrix and the correct 5PM behavior at infinity; and since the counter terms are smooth functions of the parameters anywhere between these two boundaries, the procedure is guaranteed to yield a smooth interpolation formula.

Let us recall the form of the singular separatrix term from Eq.~(\ref{E^pm_separatrix}), 
\begin{align}\label{E^pm_sing}
    E^\pm_{\rm \rm sing}(v,j) =& B^\pm(v)\log\left(\frac{\delta j}{j}\right),
\end{align}
where 
\begin{equation}
    B^\pm(v)= -\frac{R^2(v)}{\sqrt{M (6M - R(v))}}{\cal F}^\pm_\circ(R(v)),
\end{equation}
and $j_c(v)$ is given explicitly in Eq.\ (\ref{Lcofv}). 
Here $R(v)$ is the whirl (periapsis) radius of the critical geodesic obtained by taking $j\to j_c(v)$ at fixed $v$; it is given explicitly in Eq.~(\ref{eq:circular_radius}).

We will explore two resummation formulas, one obtained from the pure PM expansion (\ref{E_PM}), and another (for $\tilde E_{\rm GW}^{+}$ only) from the mixed 5PM/$\rm 3'PN$ expansion (\ref{mixed}). We denote the corresponding resummed quantities by $\tilde E^\pm_{\rm PM}$ and $\tilde E^+_{\rm mixed}$. They are constructed through
\begin{equation}\label{resum}
\tilde E^{\pm}_{\rm X} = E^{\pm}_{\rm X} +E^\pm_{\rm sing} - {\rm CT}^\pm_{\rm X},
\end{equation}
where ${\rm X}\in\{{\rm PM},{\rm mixed}\}$, $E^{\pm}_{\rm X}$ is the base PM or mixed expansion,
and ${\rm CT}^\pm_{\rm X}$ are suitable counter-terms. The latter are given by
\begin{align}
    {\rm CT}^{\pm}_{\rm PM}= -B^\pm(v)\sum_{k=1}^{\bar k^\pm} \frac{1}{k}\left(\frac{j_c(v)}{j}\right)^k
\end{align}
with $\bar k^+=5$ and $\bar k^-=7$; and
\begin{align}\label{CT_mixed}
    {\rm CT}^{+}_{\rm mixed}= {\rm CT}^{+}_{\rm PM} + \sum_{n=0}^{13} \sum_{k=6}^{15} {\rm ct}^{+}_{kn} \frac{p_{\infty}^{n}}{(p_{\infty} j)^k} \, ,
\end{align}
where in the last expression the term ${\rm CT}^{+}_{\rm PM}$ accounts for all PM expansion terms of $E^+_{\rm sing}$ up to 5PM order, and the remaining terms on the right-hand side capture the $\leq$3PN pieces of higher PM terms up to 15PM order. The sum over $n$ here is associated with the PN expansion in powers $p_\infty$ at fixed $j p_\infty$ (equivalent to fixed $e_N$), with the coefficients given by 
\begin{align}
    {\rm ct}^{+}_{kn}&= -\frac{1}{n!} \frac{\partial^n}{\partial p_{\infty}^n} \left[ B^+(v) \frac{(j_c(v))^k}{k} p_\infty^k \right] \Bigg|_{p_{\infty} = 0}\,,
\end{align}
where one makes the replacement $v=p_\infty(1+p_\infty^2)^{-1/2}$ before taking derivatives.  We note that ${\rm ct}^{+}_{kn}=0$ for $0\leq n\leq 5$ (since $k\geq 6$), and, in general, ${\rm ct}^{+}_{kn}\ne 0$ for $n=6$. Thus we may also write Eq.\ (\ref{CT_mixed}) in the somewhat more transparent form  
\begin{align}\label{CT_mixed_alt}
    {\rm CT}^{+}_{\rm mixed}= {\rm CT}^{+}_{\rm PM} + p_\infty^6 \sum_{n=0}^{7} \sum_{k=6}^{15} {\rm ct}^{+}_{k,n+6} \frac{p_{\infty}^{n}}{(p_{\infty} j)^k} \, ,
\end{align}
showing that the PN expansion of $E^+_{\rm sing}$ begins at 3PN order [half an order lower than that of $E^+_{\rm GW}$; cf.~Eq.~(\ref{E_PN})].

The numerical values of the coefficients ${\rm ct}^\pm_{kn}$ are available in the accompanying Mathematica script \cite{ZenodoMma}.

\section{Test of resummation formulas} \label{Sec:results}

In this section we illustrate the performance of our resummation formulas using  benchmark numerical data from the code of Ref.\ \cite{Warburton:2025ymy}, which are available in the accompanying ancillary files \cite{ZenodoRad,ZenodoAbs}. We have checked and confirmed that error bars on the numerical data are visibly indiscernible on the scale of the plots to be presented below (with the exception of a few data points in Fig.~\ref{fig:test_out_0.7}; see the discussion in the caption of that figure). For our purposes we thus usually regard the numerical data as ``exact''. 

\begin{figure}[tb]
    \centering
    \includegraphics[width=\linewidth]{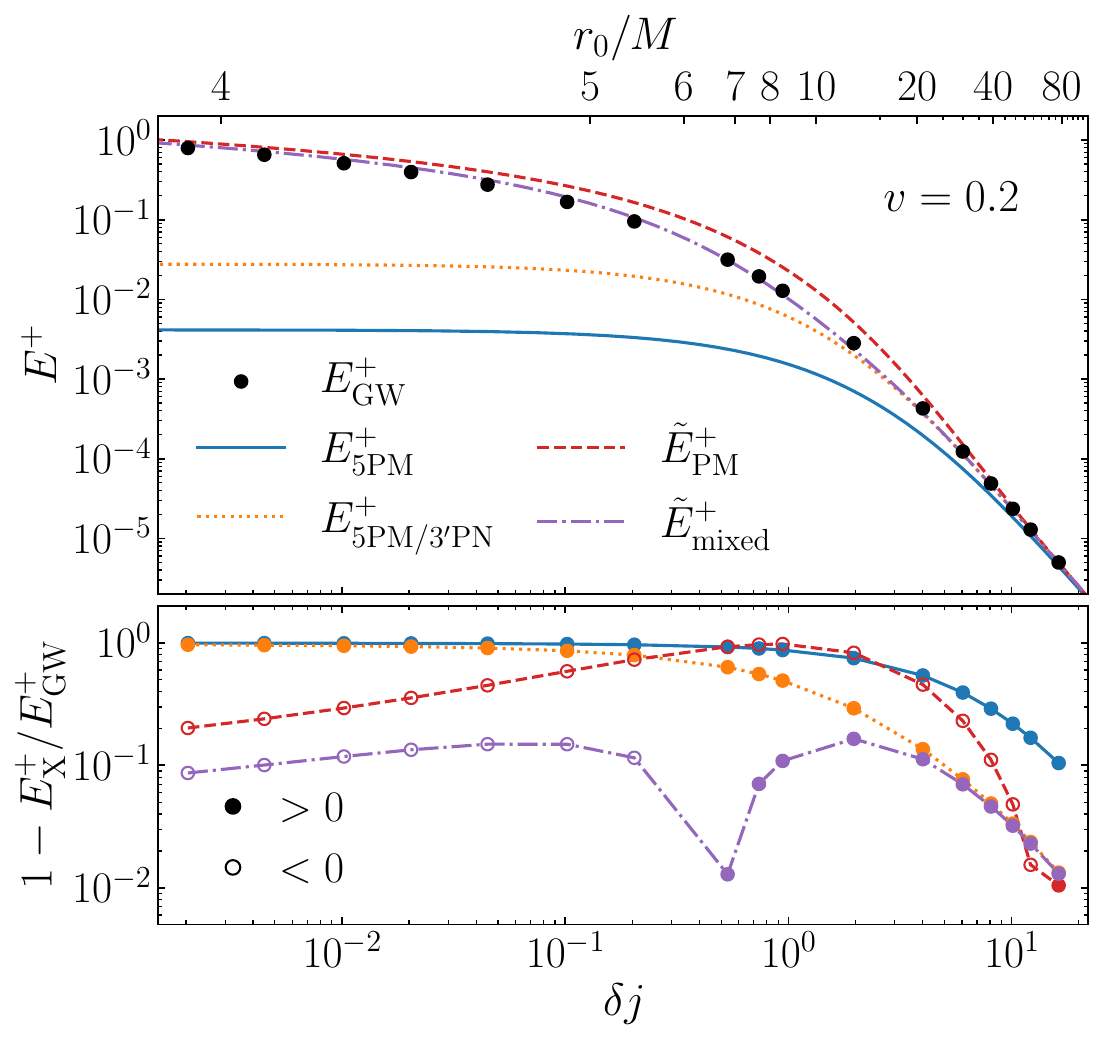}
    \caption{Resummed energy-loss models $\tilde E^+_{\rm PM}$ and $\tilde E^+_{\rm mixed}$ [Eq.~(\ref{resum})] as functions of $\delta j:=j-j_c(v)$ (lower axis) and periastron distance $r_0$ (upper axis) for fixed $v=0.2$. The resummation models are plotted against reference numerical data for $E^+_{\rm GW}$ from the code of \cite{Warburton:2025ymy}. (Estimated error bars on the numerical values are too small to be visible on the scale of this plot.). Also shown for reference are the plain 5PM and 5PM/$\rm 3'PN$ weak-field approximations (without resummation).  The lower panel displays relative differences with respect to the numerical data, with filled (open) markers denoting underestimation (overestimation) by the model.}
    \label{fig:test_out_0.2}
\end{figure}

Figure \ref{fig:test_out_0.2} displays the resummed energy-loss expressions $\tilde E^+_{\rm PM}$ and $\tilde E^+_{\rm mixed}$ as functions of $\delta j:=j-j_c(v)$ for a fixed velocity-at-infinity value of $v=0.2$. Also shown, for reference, are the plain 5PM and 5PM/$\rm 3'PN$ approximations (without resummation), and the comparison numerical results for a sample of $\delta j$ values.  By design, the resummation formulas both agree with the exact values $E^+_{\rm GW}$ in the asymptotic domains $\delta j\to 0$ and $\delta j\to \infty$, but we see in this example how they succeed in tracking the correct behavior uniformly well at any $\delta j$. While the plain weak-field approximations get progressively worse at small $\delta j$ (being at least a factor 2 off already at $r_0\sim 8M$), $\tilde E^+_{\rm mixed}$ agrees with $E^+_{\rm GW}$ to within at least $\sim 10\%$ throughout the entire domain. Notably, the resummation formulas appear to outperform the plain weak-field expressions anywhere in the domain, and even in the weak-field regime. 

\begin{figure}[thb]
    \centering
    \includegraphics[width=\linewidth]{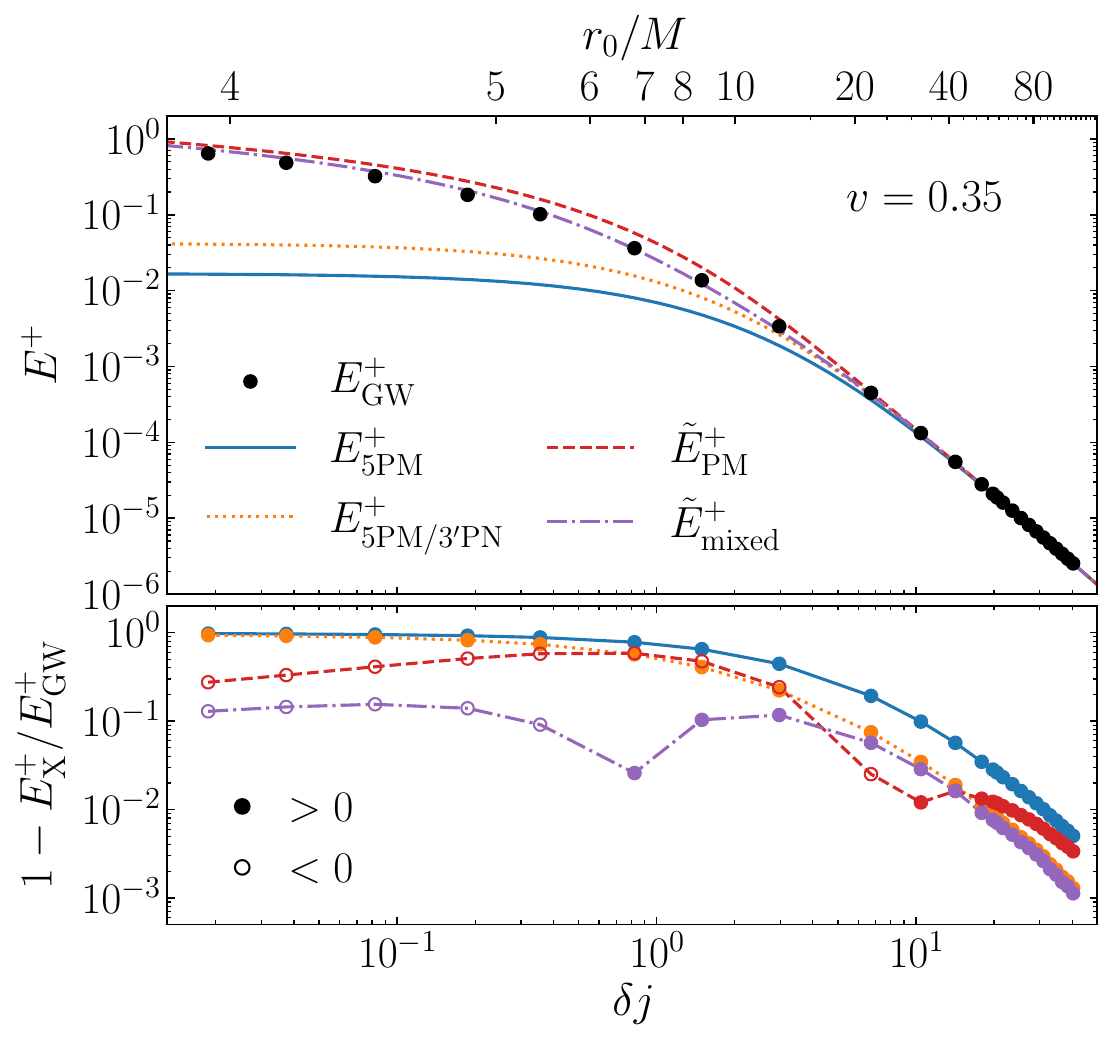}
    \caption{Same as in Fig.~\ref{fig:test_out_0.2} but for $v=0.35$.}
    \label{fig:test_out_0.35}
\end{figure}

\begin{figure}[thb]
    \centering
    \includegraphics[width=\linewidth]{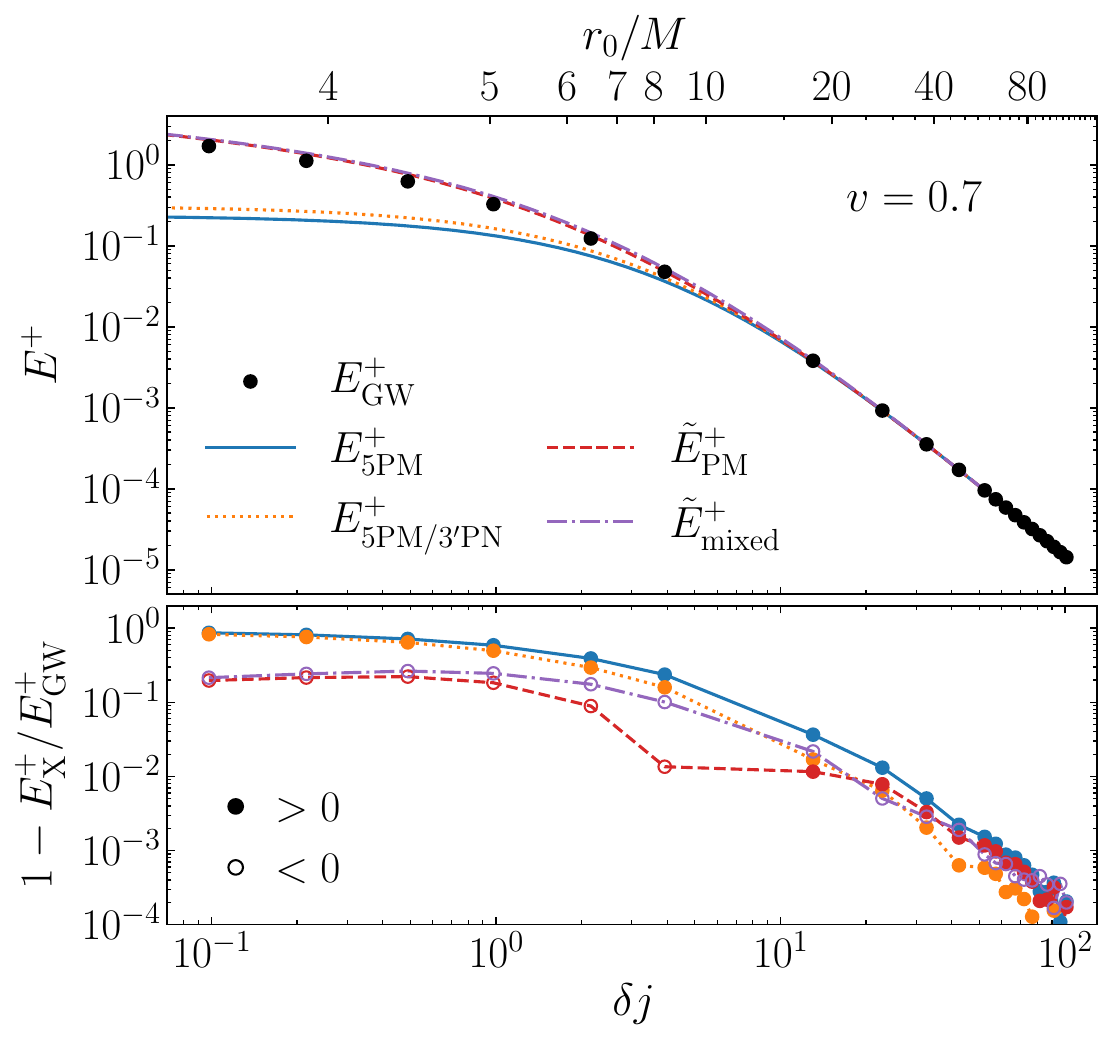}
    \caption{Same as in Figs.~\ref{fig:test_out_0.2} and \ref{fig:test_out_0.35} but for $v=0.7$. The onset of noisy behavior in the relative differences at large $\delta j$ is where differences dip below the floor of numerical error in the data. The level at which this occurs is consistent with our estimate of that numerical error. }
    \label{fig:test_out_0.7}
\end{figure}

Similar comparisons are presented in Figs.~\ref{fig:test_out_0.35} and \ref{fig:test_out_0.7} for fixed $v=0.35$ and fixed $v=0.7$, respectively. The same general behavior is observed here too: both resummation formulas describe the flux reasonably well across the entire domain, with $\tilde E^+_{\rm mixed}$ agreeing with the numerical data to within $\lesssim 15\%$ at $v=0.35$, or within $\lesssim 25\%$ at $v=0.7$. Visibly, as might be expected, the added value from including PN terms in our weak-field approximation diminishes with increasing $v$; at $v=0.7$ the resummed PM model on its own performs essentially as well.   

\begin{figure}[thb]
    \centering
    \includegraphics[width=\linewidth]{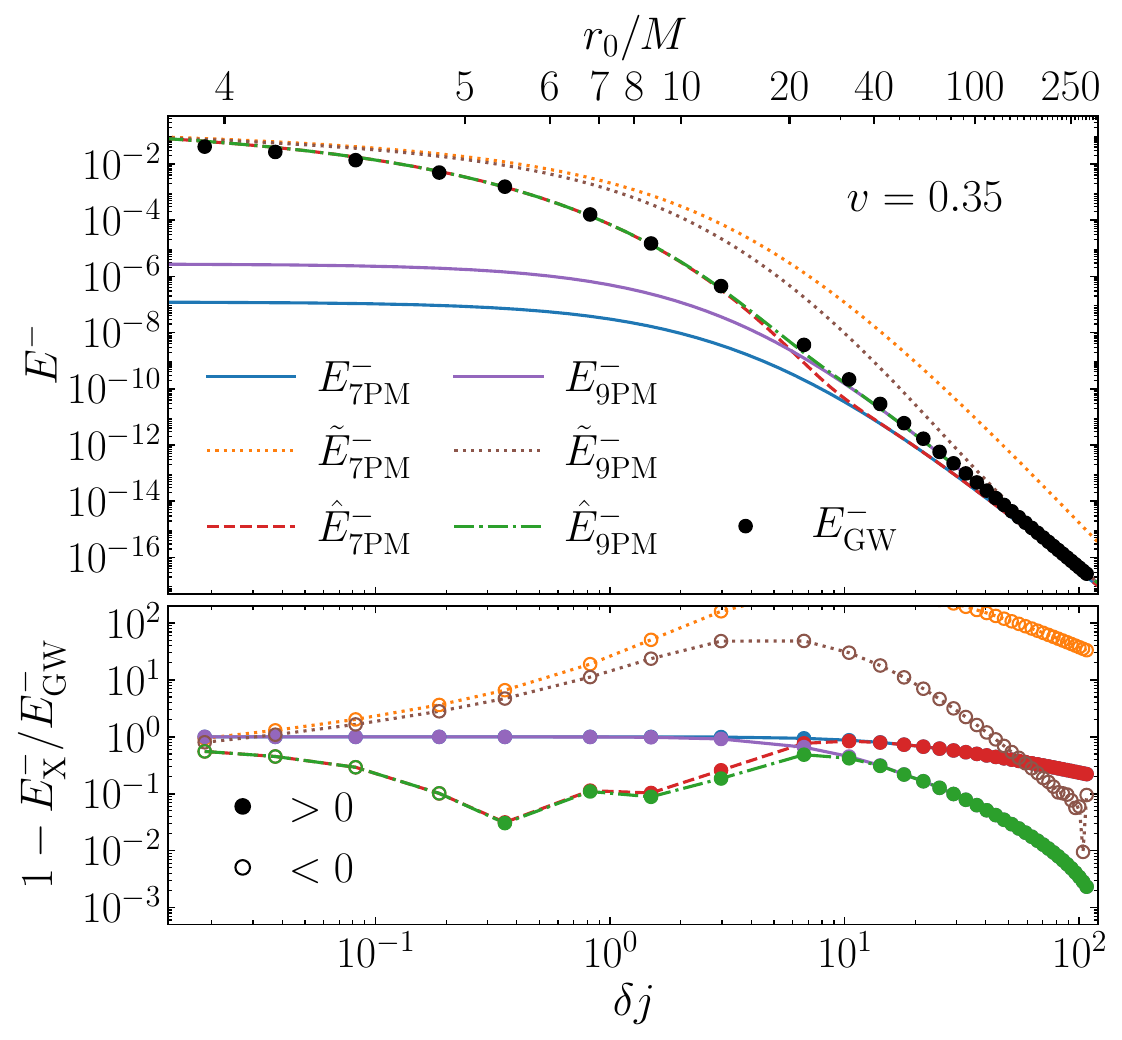}
    \caption{Resummed horizon absorption models $\tilde E^-_{\rm PM}$ [Eq.~(\ref{resum})] and $\hat E^-_{\rm PM}$ [Eq.~(\ref{resum2})], with a comparison to the plain 7PM formula [Eqs.~(\ref{E_PM}) with (\ref{E7})] and to numerical data from the code of \cite{Warburton:2025ymy}, for $v=0.35$. Also shown are the equivalent models using the 8PM and 9PM coefficients extracted from weak-field numerical data \cite{Warburton:2025ymy}. The structure of the plot is similar to that of Figs.~\ref{fig:test_out_0.2}--\ref{fig:test_out_0.7}.}
    \label{fig:test_in_0.35}
\end{figure}

We now turn to the horizon energy absorption $E^-_{\rm GW}$. Figure \ref{fig:test_in_0.35} shows the results in the example of $v=0.35$. Once again, the resummation formula, $\tilde E^-_{\rm PM}$ in this case, is plotted against the plain PM approximation, here $E^-_{\rm 7PM}$, and compared with ``exact'' numerical data from the code of \cite{Warburton:2025ymy}, which are available in the accompanying ancillary file \cite{ZenodoAbs}. We note immediately that,
in this case, the term $(E^-_{\rm sing}-{\rm CT}^-_{\rm PM})$ in Eq.~(\ref{resum}) remains much greater than $E^-_{\rm PM}$ even at large $j$, disrupting the ability of $\tilde E^-_{\rm PM}$ to capture the correct PM behavior in the weak-field regime. 

To address this problem, we are forced to abandon our usual resummation technique and introduce an alternative procedure in its place. Instead of eliminating a finite number of PM expansion terms of the singular function $E^-_{\rm sing}$ (via the subtraction of ${\rm CT}^-_{\rm PM}$), we suppress the support of $E^-_{\rm sing}$ at large $j$ more sharply with an exponential attenuation function. Specifically, in the resummation formula (\ref{resum}) we multiply $E^-_{\rm sing}$ by a function of the form $\exp[-k (\delta j/j_c)^\beta]$, with certain $k>0$ and $\beta>0$, and drop the counter-term ${\rm CT}^-_{\rm PM}$ altogether. We have found through experiment that an attenuation with $\beta=1$ acts too aggressively, producing too sharp a transition between $E^-_{\rm sing}$ and the PM approximation. The value $\beta=1/2$ was selected, as it was observed to facilitate a smooth transition. Our modified resummation formula for the absorbed energy thus takes the form   
\begin{equation}\label{resum2}
\hat E^{-}_{\rm PM} = E^-_{\rm PM} + e^{-k \sqrt{\delta j/j_c}} E^{-}_{\rm sing},
\end{equation}
where it only remains for us to prescribe the value of $k$. 

That we do using the following procedure. First, for each fixed value of $v$, we determine the value $j=j_0$ that minimizes the difference $\big|E^-_{\rm PM}(j) - E^{-}_{\rm sing}(j)\big|$; it can be easily confirmed (analytically) that this difference always admits one and only one minimum, for any value of $v$.  We then choose $j_0$ to be the value around which to introduce the transition from the $E^{-}_{\rm sing}$ term to the $E^-_{\rm PM}$ term, achieved by requiring that the $E^{-}_{\rm sing}$ term is suppressed by full 3 exponential folds (i.e., a factor $e^{-3}$) at $j=j_0$. This condition is enforced with the choice
\begin{equation}\label{resum3}
k=3\sqrt{j_c/\delta j_0},
\end{equation}
where $\delta j_0:=j_0-j_c$. This completes our prescription for the modified resummation model $\hat E^{-}_{\rm PM}$. For our specific $v=0.35$ example, we have calculated and used $k=13.8894$. 
We have plotted the alternative resummation formula $\hat E^{-}_{\rm PM}$ too in Figure \ref{fig:test_in_0.35}, showing it does much better than $\tilde E^{-}_{\rm PM}$ in providing a faithful representation of the true function $E^-_{{\rm GW}}$ across the domain.

A comparison of Figs.~\ref{fig:test_out_0.35} and \ref{fig:test_in_0.35} shows that the PM approximation $E^-_{\rm 7PM}$ captures the large-$j$ asymptotic behavior of $E^-_{\rm GW}$ less well than $E^+_{\rm 5PM}$ captures $E^+_{\rm GW}$. This might be expected from the fact that $E^-_{\rm 7PM}$ is only a leading-order approximation while $E^+_{\rm 5PM}$ is accurate to third relative order. The differing level of performance is evident in the above examples. We do not yet have a more accurate model for $E^-_{\rm GW}$, but we can attempt to ``predict'' the amplitude of its $>7{\rm PM}$ terms at given $v$ values by fitting a higher-order PM model to a large-$j$ segment of the numerical data. NW previously applied this procedure to numerically determine the coefficients of the 8PM and 9PM terms of $E^-_{\rm GW}$ for $v=0.35$ \cite{Warburton:2025ymy}. The resulting 9PM-accurate model is also plotted in Fig.~\ref{fig:test_in_0.35}, together with the corresponding resummation formulas $\tilde E^{-}_{\rm 9PM}$ and $\hat E^{-}_{\rm 9PM}$, where for the term $E^{-}_{\rm PM}$ we use the fitted 9PM model. An improved performance of the resummation models is evident in the entire domain. This experiment should serve to further motivate an analytical PM calculation of the horizon absorption beyond the leading order, which to the best of our knowledge has not been attempted so far.

%----------------------------------------------------------------------
\section{Concluding remarks and future directions}\label{sec:conclusion}
%----------------------------------------------------------------------

This work advances the idea that the universal critical behavior near the separatrix between bound and plunging orbits around black holes can serve as a strong-field diagnostic informing an effective resummation of weak-field analytical expansions. The basic idea has been explored before as applied to the scattering angle in hyperbolic encounters \cite{Damour:2022ybd,Rettegno:2023ghr,Long:2024ltn}. Here, for the first time, we apply it to radiative observables, focusing on the total energy loss in hyperbolic encounters as a first example. 

Our resummation procedure is based on the observation that, in the separatrix limit, the energy loss is dominated by radiation from a circular whirl motion at the periapsis distance, which diverges in proportion to the number of whirl cycles. The logarithmic form of this divergence is universal, and its overall scaling is determinable from the constant flux of energy emitted during the whirl motion. We have produced an accurate analytical model for these fluxes by fitting to high-precision numerical calculations for unstable circular geodesics. Hence we have produced an accurate analytical model for the divergent behavior of the energy loss. 

We have then prescribed a procedure that blends our near-separatrix model with state-of-the-art weak-field models for the $O(q^2)$ energy loss to infinity (accurate to 5PM and 3PN orders) and down the event horizon (accurate to 7PM order). The resulting formulas effectively interpolate between the weak and strong-field regimes, realizing the synergistic potential between the PM, PN and self-force methods. The outcome is a semi-analytical approximant that remains uniformly reliable across the full scattering parameter space. We have demonstrated the performance of our resummed energy-loss formulas against accurate results from numerical calculations for scattering orbits. 

As we have alluded to at the start of Sec.~\ref{Sec:critical}, the separatrix singularity is but an artifact of ignoring radiation-reaction correction to the orbit; of course, no divergence occurs in reality. Generically, radiative dynamics transports the object across the separatrix on a timescale proportional to $\log q$ \cite{OShaughnessy:2002tbu}, after which the object plunges into the black hole on a near-geodesic trajectory. (In \cite{Gundlach:2012aj} it was shown that, via exponential fine-tuning of the initial conditions, one can arrange for the particle to maintain a whirl motion near the separatrix for as long as a full radiation timescale, $\sim 1/q$.) Infinite whirl motion formally occurs only in the limit $q\to 0$. We therefore expect our method to work well only for sufficiently small mass ratios, and better with decreasing $q$.   

We expect the basic idea of our procedure to be applicable to other scattering observables, including the angular-momentum loss, the gravitational time-delay, or the gravitational redshift along scattering orbits. More profoundly, the procedure should offer a systematic route to constructing uniformly valid models even for {\it bound-orbit} observables, either via scattering-to-bound mappings or through a direct resummation of PN fluxes informed by bound-orbit separatrix physics. More broadly, this approach points toward a new organizing principle for gravitational-wave observables, in which analyticity and critical behavior play a central role in bridging between the weak-field and strong-field regimes.

%----------------------------------------------------------------------
\textit{Acknowledgments---} We thank David Trestini, Zach Nasipak, and Adam Pound for useful discussions, and Mark Hurrell for bringing to our attention several typos in the original version of this manuscript. We are grateful to the participants of the ``2nd Annual Workshop on Self-Force and Amplitudes'' in Southampton, where this work was presented for the first time, for stimulating discussions. 
L.B. acknowledges support from the STFC via Grant No.~ST/B001170/1.
The work of R.G.\ is supported by the Royal Society grant RF\textbackslash ERE\textbackslash 231084. 
B.L. gratefully acknowledges funding from the European Union’s Horizon Europe research and innovation programme under the Marie Skłodowska-Curie grant agreement No.~101209791. 
N.W. acknowledges
support from a Royal Society – Research Ireland University Research Fellowship. 
This publication has emanated from research conducted with the financial support of Research Ireland under grant number 22/RS-URF-R/3825. 
Funded by the European Union (ERC grant GWSky/101167314). Views and opinions expressed are however those of the author(s) only and do not necessarily reflect those of the European Union or the European Research Council Executive Agency. Neither the European Union nor the granting authority can be held responsible for them.
We acknowledge the use of the IRIDIS High Performance Computing Facility, and associated support services at the University of Southampton, in the completion of this work. This work makes use of the Black Hole Perturbation Toolkit \cite{BHPToolkit}.

\appendix

%----------------------------------------------------------------------

%----------------------------------------------------------------------

\newpage

\bibliography{references}% Produces the bibliography via BibTeX.

@misc{BHPToolkit,
  title = {{Black Hole Perturbation Toolkit}},
  howpublished = {(\href{http://bhptoolkit.org/}{bhptoolkit.org})},
}

@article{Adamo:2024oxy,
    author = "Adamo, Tim and Gonzo, Riccardo and Ilderton, Anton",
    title = "{Gravitational bound waveforms from amplitudes}",
    eprint = "2402.00124",
    archivePrefix = "arXiv",
    primaryClass = "hep-th",
    doi = "10.1007/JHEP05(2024)034",
    journal = "JHEP",
    volume = "05",
    pages = "034",
    year = "2024"
}

@article{Antonelli:2019ytb,
    author = "Antonelli, Andrea and Buonanno, Alessandra and Steinhoff, Jan and van de Meent, Maarten and Vines, Justin",
    title = "{Energetics of two-body Hamiltonians in post-Minkowskian gravity}",
    eprint = "1901.07102",
    archivePrefix = "arXiv",
    primaryClass = "gr-qc",
    doi = "10.1103/PhysRevD.99.104004",
    journal = "Phys. Rev. D",
    volume = "99",
    number = "10",
    pages = "104004",
    year = "2019"
}

@article{Akcay:2012ea,
    author = "Akcay, Sarp and Barack, Leor and Damour, Thibault and Sago, Norichika",
    title = "{Gravitational self-force and the effective-one-body formalism between the innermost stable circular orbit and the light ring}",
    eprint = "1209.0964",
    archivePrefix = "arXiv",
    primaryClass = "gr-qc",
    doi = "10.1103/PhysRevD.86.104041",
    journal = "Phys. Rev. D",
    volume = "86",
    pages = "104041",
    year = "2012"
}

@article{Akpinar:2025huz,
    author = "Akpinar, Dogan and del Duca, Vittorio and Gonzo, Riccardo",
    title = "{The spinning self-force EFT: 1SF waveform recursion relation and Compton scattering}",
    eprint = "2504.02025",
    archivePrefix = "arXiv",
    journal = "",
    primaryClass = "hep-th",
    month = "4",
    year = "2025"
}

@article{Barack:2019agd,
    author = "Barack, Leor and Colleoni, Marta and Damour, Thibault and Isoyama, Soichiro and Sago, Norichika",
    title = "{Self-force effects on the marginally bound zoom-whirl orbit in Schwarzschild spacetime}",
    eprint = "1909.06103",
    archivePrefix = "arXiv",
    primaryClass = "gr-qc",
    doi = "10.1103/PhysRevD.100.124015",
    journal = "Phys. Rev. D",
    volume = "100",
    number = "12",
    pages = "124015",
    year = "2019"
}

@article{Barausse:2021xes,
    author = "Barausse, Enrico and Berti, Emanuele and Cardoso, Vitor and Hughes, Scott A. and Khanna, Gaurav",
    title = "{Divergences in gravitational-wave emission and absorption from extreme mass ratio binaries}",
    eprint = "2106.09721",
    archivePrefix = "arXiv",
    primaryClass = "gr-qc",
    doi = "10.1103/PhysRevD.104.064031",
    journal = "Phys. Rev. D",
    volume = "104",
    number = "6",
    pages = "064031",
    year = "2021"
}

@article{Berti:2010ce,
    author = "Berti, Emanuele and Cardoso, Vitor and Hinderer, Tanja and Lemos, Madalena and Pretorius, Frans and Sperhake, Ulrich and Yunes, Nicolas",
    title = "{Semianalytical estimates of scattering thresholds and gravitational radiation in ultrarelativistic black hole encounters}",
    eprint = "1003.0812",
    archivePrefix = "arXiv",
    primaryClass = "gr-qc",
    doi = "10.1103/PhysRevD.81.104048",
    journal = "Phys. Rev. D",
    volume = "81",
    pages = "104048",
    year = "2010"
}

@article{Bini:2021gat,
    author = "Bini, Donato and Damour, Thibault and Geralico, Andrea",
    title = "{Radiative contributions to gravitational scattering}",
    eprint = "2107.08896",
    archivePrefix = "arXiv",
    primaryClass = "gr-qc",
    doi = "10.1103/PhysRevD.104.084031",
    journal = "Phys. Rev. D",
    volume = "104",
    number = "8",
    pages = "084031",
    year = "2021"
}

@article{Goldberger:2005cd,
    author = "Goldberger, Walter D. and Rothstein, Ira Z.",
    title = "{Dissipative effects in the worldline approach to black hole dynamics}",
    eprint = "hep-th/0511133",
    archivePrefix = "arXiv",
    doi = "10.1103/PhysRevD.73.104030",
    journal = "Phys. Rev. D",
    volume = "73",
    pages = "104030",
    year = "2006"
}

@article{Bern:2025wyd,
    author = "Bern, Zvi and Herrmann, Enrico and Roiban, Radu and Ruf, Michael S. and Smirnov, Alexander V. and Smith, Sid and Zeng, Mao",
    title = "{Scattering Amplitudes and Conservative Binary Dynamics at $O(G^5)$ without Self-Force Truncation}",
    eprint = "2512.23654",
    archivePrefix = "arXiv",
    primaryClass = "hep-th",
    journal ="",
    month = "12",
    year = "2025"
}

@article{Bini:2022enm,
    author = "Bini, Donato and Damour, Thibault and Geralico, Andrea",
    title = "{Radiated momentum and radiation reaction in gravitational two-body scattering including time-asymmetric effects}",
    eprint = "2210.07165",
    archivePrefix = "arXiv",
    primaryClass = "gr-qc",
    doi = "10.1103/PhysRevD.107.024012",
    journal = "Phys. Rev. D",
    volume = "107",
    number = "2",
    pages = "024012",
    year = "2023"
}

@article{Blanchet:1992br,
    author = "Blanchet, Luc and Damour, Thibault",
    title = "{Hereditary effects in gravitational radiation}",
    reportNumber = "IHES-P-92-42",
    doi = "10.1103/PhysRevD.46.4304",
    journal = "Phys. Rev. D",
    volume = "46",
    pages = "4304--4319",
    year = "1992"
}

@article{Blanchet:2013haa,
    author = "Blanchet, Luc",
    title = "{Post-Newtonian Theory for Gravitational Waves}",
    eprint = "1310.1528",
    archivePrefix = "arXiv",
    primaryClass = "gr-qc",
    doi = "10.12942/lrr-2014-2",
    journal = "Living Rev. Rel.",
    volume = "17",
    pages = "2",
    year = "2014"
}

@article{Buonanno:2024vkx,
    author = "Buonanno, Alessandra and Jakobsen, Gustav Uhre and Mogull, Gustav",
    title = "{Post-Minkowskian theory meets the spinning effective-one-body approach for two-body scattering}",
    eprint = "2402.12342",
    archivePrefix = "arXiv",
    primaryClass = "gr-qc",
    reportNumber = "HU-EP-24/07-RTG",
    doi = "10.1103/PhysRevD.110.044038",
    journal = "Phys. Rev. D",
    volume = "110",
    number = "4",
    pages = "044038",
    year = "2024"
}

@article{Buonanno:2024byg,
	archiveprefix = {arXiv},
	author = {Buonanno, Alessandra and Mogull, Gustav and Patil, Raj and Pompili, Lorenzo},
	doi = {10.1103/PhysRevLett.133.211402},
	eprint = {2405.19181},
	journal = {Phys. Rev. Lett.},
	number = {21},
	pages = {211402},
	primaryclass = {gr-qc},
	reportnumber = {HU-EP-24/16-RTG},
	title = {{Post-Minkowskian Theory Meets the Spinning Effective-One-Body Approach for Bound-Orbit Waveforms}},
	volume = {133},
	year = {2024}}

@article{Cho:2021onr,
    author = "Cho, Gihyuk and Dandapat, Subhajit and Gopakumar, Achamveedu",
    title = "{Third order post-Newtonian gravitational radiation from two-body scattering: Instantaneous energy and angular momentum radiation}",
    eprint = "2111.00818",
    archivePrefix = "arXiv",
    primaryClass = "gr-qc",
    doi = "10.1103/PhysRevD.105.084018",
    journal = "Phys. Rev. D",
    volume = "105",
    number = "8",
    pages = "084018",
    year = "2022"
}

@article{Cho:2022pqy,
    author = "Cho, Gihyuk",
    title = "{Third post-Newtonian gravitational radiation from two-body scattering. II. Hereditary energy radiation}",
    eprint = "2203.10872",
    archivePrefix = "arXiv",
    primaryClass = "gr-qc",
    doi = "10.1103/PhysRevD.105.104035",
    journal = "Phys. Rev. D",
    volume = "105",
    number = "10",
    pages = "104035",
    year = "2022"
}

@article{Cho:2021arx,
    author = {Cho, Gihyuk and K{\"a}lin, Gregor and Porto, Rafael A.},
    title = "{From boundary data to bound states. Part III. Radiative effects}",
    eprint = "2112.03976",
    archivePrefix = "arXiv",
    primaryClass = "hep-th",
    reportNumber = "DESY 21-212",
    doi = "10.1007/JHEP04(2022)154",
    journal = "JHEP",
    volume = "04",
    pages = "154",
    year = "2022",
    note = "[Erratum: JHEP 07, 002 (2022)]"
}

@article{Cipriani:2025ikx,
    author = "Cipriani, Andrea and Di Russo, Giorgio and Fucito, Francesco and Morales, Jos{\'e} Francisco and Poghosyan, Hasmik and Poghossian, Rubik",
    title = "{Resumming Post-Minkowskian and Post-Newtonian gravitational waveform expansions}",
    eprint = "2501.19257",
    journal = "",
    archivePrefix = "arXiv",
    primaryClass = "gr-qc",
    month = "1",
    year = "2025"
}

@article{Cutler:1994,
  title = {Gravitational radiation reaction for bound motion around a Schwarzschild black hole},
  author = {Cutler, Curt and Kennefick, Daniel and Poisson, Eric},
  journal = {Phys. Rev. D},
  volume = {50},
  issue = {6},
  pages = {3816--3835},
  numpages = {0},
  year = {1994},
  month = {Sep},
  publisher = {American Physical Society},
  doi = {10.1103/PhysRevD.50.3816},
  url = {https://link.aps.org/doi/10.1103/PhysRevD.50.3816}
}

@article{Damgaard:2023ttc,
    author = "Damgaard, Poul H. and Hansen, Elias Roos and Plant{\'e}, Ludovic and Vanhove, Pierre",
    title = "{Classical observables from the exponential representation of the gravitational S-matrix}",
    eprint = "2307.04746",
    archivePrefix = "arXiv",
    primaryClass = "hep-th",
    reportNumber = "CERN-TH-2023-135, IPhT-T23/041, LAPTh-029/23",
    doi = "10.1007/JHEP09(2023)183",
    journal = "JHEP",
    volume = "09",
    pages = "183",
    year = "2023"
}

@article{Damour:1988mr,
    author = "Damour, T. and Schaefer, Gerhard",
    title = "{Higher Order Relativistic Periastron Advances and Binary Pulsars}",
    reportNumber = "MEUDON-88024",
    doi = "10.1007/BF02828697",
    journal = "Nuovo Cim. B",
    volume = "101",
    pages = "127",
    year = "1988"
}

@article{Damour:2020tta,
    author = "Damour, Thibault",
    title = "{Radiative contribution to classical gravitational scattering at the third order in $G$}",
    eprint = "2010.01641",
    archivePrefix = "arXiv",
    primaryClass = "gr-qc",
    doi = "10.1103/PhysRevD.102.124008",
    journal = "Phys. Rev. D",
    volume = "102",
    number = "12",
    pages = "124008",
    year = "2020"
}

@article{Damour:2022ybd,
    author = "Damour, Thibault and Rettegno, Piero",
    title = "{Strong-field scattering of two black holes: Numerical relativity meets post-Minkowskian gravity}",
    eprint = "2211.01399",
    archivePrefix = "arXiv",
    primaryClass = "gr-qc",
    doi = "10.1103/PhysRevD.107.064051",
    journal = "Phys. Rev. D",
    volume = "107",
    number = "6",
    pages = "064051",
    year = "2023"
}

@article{DiVecchia:2022piu,
    author = "Di Vecchia, Paolo and Heissenberg, Carlo and Russo, Rodolfo and Veneziano, Gabriele",
    title = "{Classical gravitational observables from the Eikonal operator}",
    eprint = "2210.12118",
    archivePrefix = "arXiv",
    primaryClass = "hep-th",
    reportNumber = "CERN-TH-2022-169, NORDITA 2022-071, QMUL-PH-22-32, UUITP-42/22",
    doi = "10.1016/j.physletb.2023.138049",
    journal = "Phys. Lett. B",
    volume = "843",
    pages = "138049",
    year = "2023"
}

@article{Dlapa:2022lmu,
    author = {Dlapa, Christoph and K{\"a}lin, Gregor and Liu, Zhengwen and Neef, Jakob and Porto, Rafael A.},
    title = "{Radiation Reaction and Gravitational Waves at Fourth Post-Minkowskian Order}",
    eprint = "2210.05541",
    archivePrefix = "arXiv",
    primaryClass = "hep-th",
    doi = "10.1103/PhysRevLett.130.101401",
    journal = "Phys. Rev. Lett.",
    volume = "130",
    number = "10",
    pages = "101401",
    year = "2023"
}

@article{Driesse:2024feo,
    author = "Driesse, Mathias and Jakobsen, Gustav Uhre and Klemm, Albrecht and Mogull, Gustav and Nega, Christoph and Plefka, Jan and Sauer, Benjamin and Usovitsch, Johann",
    title = "{Emergence of Calabi{\textendash}Yau manifolds in high-precision black-hole scattering}",
    eprint = "2411.11846",
    archivePrefix = "arXiv",
    primaryClass = "hep-th",
    reportNumber = "HU-EP-24/32-RTG, QMUL-PH-24-26, BONN-TH-2024-15, TUM-HEP-1532/24",
    doi = "10.1038/s41586-025-08984-2",
    journal = "Nature",
    volume = "641",
    number = "8063",
    pages = "603--607",
    year = "2025"
}

@article{Cipriani:2026myb,
    author = "Cipriani, Andrea and Fucito, Francesco and Heissenberg, Carlo and Morales, Jose Francisco and Russo, Rodolfo",
    title = "{''Waveforms'' at the Horizon}",
    eprint = "2602.05766",
    journal = "",
    archivePrefix = "arXiv",
    primaryClass = "gr-qc",
    month = "2",
    year = "2026"
}

@article{Driesse:2026qiz,
    author = "Driesse, Mathias and Jakobsen, Gustav Uhre and Mogull, Gustav and Nega, Christoph and Plefka, Jan and Sauer, Benjamin and Usovitsch, Johann",
    title = "{Conservative Black Hole Scattering at Fifth Post-Minkowskian and Second Self-Force Order}",
    eprint = "2601.16256",
    archivePrefix = "arXiv",
    primaryClass = "hep-th",
    reportNumber = "HU-EP-26/04-RTG",
    journal ="",
    month = "1",
    year = "2026"
}

@article{Fujita:2012cm,
    author = "Fujita, Ryuichi",
    title = "{Gravitational Waves from a Particle in Circular Orbits around a Schwarzschild Black Hole to the 22nd Post-Newtonian Order}",
    eprint = "1211.5535",
    archivePrefix = "arXiv",
    primaryClass = "gr-qc",
    doi = "10.1143/PTP.128.971",
    journal = "Prog. Theor. Phys.",
    volume = "128",
    pages = "971--992",
    year = "2012"
}

@article{Fujita:2014eta,
    author = "Fujita, Ryuichi",
    title = "{Gravitational Waves from a Particle in Circular Orbits around a Rotating Black Hole to the 11th Post-Newtonian Order}",
    eprint = "1412.5689",
    archivePrefix = "arXiv",
    primaryClass = "gr-qc",
    doi = "10.1093/ptep/ptv012",
    journal = "PTEP",
    volume = "2015",
    number = "3",
    pages = "033E01",
    year = "2015"
}

@article{Glampedakis:2002ya,
    author = "Glampedakis, Kostas and Kennefick, Daniel",
    title = "{Zoom and whirl: Eccentric equatorial orbits around spinning black holes and their evolution under gravitational radiation reaction}",
    eprint = "gr-qc/0203086",
    archivePrefix = "arXiv",
    doi = "10.1103/PhysRevD.66.044002",
    journal = "Phys. Rev. D",
    volume = "66",
    pages = "044002",
    year = "2002"
}

@article{Goldberger:2004jt,
    author = "Goldberger, Walter D. and Rothstein, Ira Z.",
    title = "{An Effective field theory of gravity for extended objects}",
    eprint = "hep-th/0409156",
    archivePrefix = "arXiv",
    reportNumber = "UCSD-PTH-04-17, CMU-HEP-04-06",
    doi = "10.1103/PhysRevD.73.104029",
    journal = "Phys. Rev. D",
    volume = "73",
    pages = "104029",
    year = "2006"
}

@article{Goldberger:2020wbx,
    author = "Goldberger, Walter D. and Rothstein, Ira Z.",
    title = "{Horizon radiation reaction forces}",
    eprint = "2007.00731",
    archivePrefix = "arXiv",
    primaryClass = "hep-th",
    doi = "10.1007/JHEP10(2020)026",
    journal = "JHEP",
    volume = "10",
    pages = "026",
    year = "2020"
}

@article{Gonzo:2023goe,
    author = "Gonzo, Riccardo and Shi, Canxin",
    title = "{Boundary to bound dictionary for generic Kerr orbits}",
    eprint = "2304.06066",
    archivePrefix = "arXiv",
    primaryClass = "hep-th",
    doi = "10.1103/PhysRevD.108.084065",
    journal = "Phys. Rev. D",
    volume = "108",
    number = "8",
    pages = "084065",
    year = "2023"
}

@article{Gonzo:2024xjk,
    author = "Gonzo, Riccardo and Lewis, Jack and Pound, Adam",
    title = "{The first law of binary black hole scattering}",
    eprint = "2409.03437",
    journal = "",
    archivePrefix = "arXiv",
    primaryClass = "gr-qc",
    month = "9",
    year = "2024"
}

@article{Gundlach:2012aj,
    author = "Gundlach, Carsten and Akcay, Sarp and Barack, Leor and Nagar, Alessandro",
    title = "{Critical phenomena at the threshold of immediate merger in binary black hole systems: the extreme mass ratio case}",
    eprint = "1207.5167",
    archivePrefix = "arXiv",
    primaryClass = "gr-qc",
    doi = "10.1103/PhysRevD.86.084022",
    journal = "Phys. Rev. D",
    volume = "86",
    pages = "084022",
    year = "2012"
}

@article{Herrmann:2021tct,
    author = "Herrmann, Enrico and Parra-Martinez, Julio and Ruf, Michael S. and Zeng, Mao",
    title = "{Radiative classical gravitational observables at $ \mathcal{O} $(G$^{3}$) from scattering amplitudes}",
    eprint = "2104.03957",
    archivePrefix = "arXiv",
    primaryClass = "hep-th",
    doi = "10.1007/JHEP10(2021)148",
    journal = "JHEP",
    volume = "10",
    pages = "148",
    year = "2021"
}

@article{Herrmann:2021lqe,
    author = "Herrmann, Enrico and Parra-Martinez, Julio and Ruf, Michael S. and Zeng, Mao",
    title = "{Gravitational Bremsstrahlung from Reverse Unitarity}",
    eprint = "2101.07255",
    archivePrefix = "arXiv",
    primaryClass = "hep-th",
    reportNumber = "CALT-TH-2021-003, FR-PHENO-2021-02, OUTP-21-02P",
    doi = "10.1103/PhysRevLett.126.201602",
    journal = "Phys. Rev. Lett.",
    volume = "126",
    number = "20",
    pages = "201602",
    year = "2021"
}

@article{Jakobsen:2023hig,
    author = "Jakobsen, Gustav Uhre and Mogull, Gustav and Plefka, Jan and Sauer, Benjamin",
    title = "{Dissipative Scattering of Spinning Black Holes at Fourth Post-Minkowskian Order}",
    eprint = "2308.11514",
    archivePrefix = "arXiv",
    primaryClass = "hep-th",
    reportNumber = "HU-EP-23/47-RTG",
    doi = "10.1103/PhysRevLett.131.241402",
    journal = "Phys. Rev. Lett.",
    volume = "131",
    number = "24",
    pages = "241402",
    year = "2023"
}

@article{Kalin:2019rwq,
    author = {K{\"a}lin, Gregor and Porto, Rafael A.},
    title = "{From Boundary Data to Bound States}",
    eprint = "1910.03008",
    archivePrefix = "arXiv",
    primaryClass = "hep-th",
    reportNumber = "DESY 19-167, UUITP-40/19, DESY-19-167",
    doi = "10.1007/JHEP01(2020)072",
    journal = "JHEP",
    volume = "01",
    pages = "072",
    year = "2020"
}

@article{Khalil:2022ylj,
    author = "Khalil, Mohammed and Buonanno, Alessandra and Steinhoff, Jan and Vines, Justin",
    title = "{Energetics and scattering of gravitational two-body systems at fourth post-Minkowskian order}",
    eprint = "2204.05047",
    archivePrefix = "arXiv",
    primaryClass = "gr-qc",
    doi = "10.1103/PhysRevD.106.024042",
    journal = "Phys. Rev. D",
    volume = "106",
    number = "2",
    pages = "024042",
    year = "2022"
}

@article{Kosower:2018adc,
    author = "Kosower, David A. and Maybee, Ben and O'Connell, Donal",
    title = "{Amplitudes, Observables, and Classical Scattering}",
    eprint = "1811.10950",
    archivePrefix = "arXiv",
    primaryClass = "hep-th",
    doi = "10.1007/JHEP02(2019)137",
    journal = "JHEP",
    volume = "02",
    pages = "137",
    year = "2019"
}

@article{Jones:2023ugm,
    author = "Jones, Callum R. T. and Ruf, Michael S.",
    title = "{Absorptive effects and classical black hole scattering}",
    eprint = "2310.00069",
    archivePrefix = "arXiv",
    primaryClass = "hep-th",
    doi = "10.1007/JHEP03(2024)015",
    journal = "JHEP",
    volume = "03",
    pages = "015",
    year = "2024"
}

@article{Leather:2024mls,
    author = "Leather, Benjamin",
    title = "{Gravitational self-force with hyperboloidal slicing and spectral methods}",
    eprint = "2411.14976",
    archivePrefix = "arXiv",
    primaryClass = "gr-qc",
    doi = "10.1007/s10714-025-03443-9",
    journal = "Gen. Rel. Grav.",
    volume = "57",
    number = "7",
    pages = "112",
    year = "2025"
}

@article{Long:2024ltn,
    author = "Long, Oliver and Whittall, Christopher and Barack, Leor",
    title = "{Black hole scattering near the transition to plunge: Self-force and resummation of post-Minkowskian theory}",
    eprint = "2406.08363",
    archivePrefix = "arXiv",
    primaryClass = "gr-qc",
    doi = "10.1103/PhysRevD.110.044039",
    journal = "Phys. Rev. D",
    volume = "110",
    number = "4",
    pages = "044039",
    year = "2024"
}

@article{Long:2025nmj,
    author = {Long, Oliver and Pfeiffer, Harald P. and Buonanno, Alessandra and Jakobsen, Gustav Uhre and Mogull, Gustav and Ramos-Buades, Antoni and R{\"u}ter, Hannes R. and Kidder, Lawrence E. and Scheel, Mark A.},
    title = "{Highly accurate simulations of asymmetric black-hole scattering and cross validation of effective-one-body models}",
    eprint = "2507.08071",
    archivePrefix = "arXiv",
    journal = "",
    primaryClass = "gr-qc",
    month = "7",
    year = "2025"
}

@article{Keldysh:1964ud,
    author = "Keldysh, L. V.",
    title = "{Diagram Technique for Nonequilibrium Processes}",
    doi = "10.1142/9789811279461_0007",
    journal = "Sov. Phys. JETP",
    volume = "20",
    pages = "1018--1026",
    year = "1965"
}

@article{Pound:2021qin,
    author = "Pound, Adam and Wardell, Barry",
    title = "{Black hole perturbation theory and gravitational self-force}",
    eprint = "2101.04592",
    archivePrefix = "arXiv",
    primaryClass = "gr-qc",
    journal = "",
    month = "1",
    year = "2021"
}

@article{Bern:2019nnu,
    author = "Bern, Zvi and Cheung, Clifford and Roiban, Radu and Shen, Chia-Hsien and Solon, Mikhail P. and Zeng, Mao",
    title = "{Scattering Amplitudes and the Conservative Hamiltonian for Binary Systems at Third Post-Minkowskian Order}",
    eprint = "1901.04424",
    archivePrefix = "arXiv",
    primaryClass = "hep-th",
    reportNumber = "CALT-TH 2019-002, UCLA/TEP/2019/101",
    doi = "10.1103/PhysRevLett.122.201603",
    journal = "Phys. Rev. Lett.",
    volume = "122",
    number = "20",
    pages = "201603",
    year = "2019"
}

@article{Hansen:1972jt,
    author = "Hansen, R. O.",
    title = "{Post-Newtonian gravitational radiation from point masses in a hyperbolic kepler orbit}",
    doi = "10.1103/PhysRevD.5.1021",
    journal = "Phys. Rev. D",
    volume = "5",
    pages = "1021--1023",
    year = "1972"
}

@article{Blanchet:1989cu,
    author = "Blanchet, Luc and Schaefer, Gerhard",
    title = "{Higher order gravitational radiation losses in binary systems}",
    journal = "Mon. Not. Roy. Astron. Soc.",
    volume = "239",
    pages = "845--867",
    year = "1989",
    note = "[Erratum: Mon.Not.Roy.Astron.Soc. 242, 704 (1990)]"
}

@article{Junker:1992kle,
    author = {Junker, Wolfgang and Sch{\"a}fer, Gerhard},
    title = "{Binary systems: higher order gravitational radiation damping and wave emission}",
    doi = "10.1093/mnras/254.1.146",
    journal = "Mon. Not. Roy. Astron. Soc.",
    volume = "254",
    number = "1",
    pages = "146--164",
    year = "1992"
}

@ARTICLE{Turner:1977,
       author = {{Turner}, M.},
        title = "{Gravitational radiation from point-masses in unbound orbits: Newtonian results.}",
      journal = {\apj},
         year = 1977,
        month = sep,
       volume = {216},
        pages = {610-619},
          doi = {10.1086/155501},
       adsurl = {https://ui.adsabs.harvard.edu/abs/1977ApJ...216..610T}
}

@article{Schwinger:1960qe,
    author = "Schwinger, Julian S.",
    title = "{Brownian motion of a quantum oscillator}",
    doi = "10.1063/1.1703727",
    journal = "J. Math. Phys.",
    volume = "2",
    pages = "407--432",
    year = "1961"
}

@article{Dlapa:2021npj,
    author = {Dlapa, Christoph and K{\"a}lin, Gregor and Liu, Zhengwen and Porto, Rafael A.},
    title = "{Dynamics of binary systems to fourth Post-Minkowskian order from the effective field theory approach}",
    eprint = "2106.08276",
    archivePrefix = "arXiv",
    primaryClass = "hep-th",
    reportNumber = "DESY 21-093, DESY-21-093, MPP-2021-83",
    doi = "10.1016/j.physletb.2022.137203",
    journal = "Phys. Lett. B",
    volume = "831",
    pages = "137203",
    year = "2022"
}

@article{Dlapa:2021vgp,
    author = {Dlapa, Christoph and K{\"a}lin, Gregor and Liu, Zhengwen and Porto, Rafael A.},
    title = "{Conservative Dynamics of Binary Systems at Fourth Post-Minkowskian Order in the Large-Eccentricity Expansion}",
    eprint = "2112.11296",
    archivePrefix = "arXiv",
    primaryClass = "hep-th",
    reportNumber = "DESY 21-226",
    doi = "10.1103/PhysRevLett.128.161104",
    journal = "Phys. Rev. Lett.",
    volume = "128",
    number = "16",
    pages = "161104",
    year = "2022"
}

@article{Bern:2021yeh,
    author = "Bern, Zvi and Parra-Martinez, Julio and Roiban, Radu and Ruf, Michael S. and Shen, Chia-Hsien and Solon, Mikhail P. and Zeng, Mao",
    title = "{Scattering Amplitudes, the Tail Effect, and Conservative Binary Dynamics at O(G4)}",
    eprint = "2112.10750",
    archivePrefix = "arXiv",
    primaryClass = "hep-th",
    doi = "10.1103/PhysRevLett.128.161103",
    journal = "Phys. Rev. Lett.",
    volume = "128",
    number = "16",
    pages = "161103",
    year = "2022"
}

@article{DiVecchia:2021bdo,
    author = "Di Vecchia, Paolo and Heissenberg, Carlo and Russo, Rodolfo and Veneziano, Gabriele",
    title = "{The eikonal approach to gravitational scattering and radiation at $ \mathcal{O} $(G$^{3}$)}",
    eprint = "2104.03256",
    archivePrefix = "arXiv",
    primaryClass = "hep-th",
    reportNumber = "CERN-TH-2021-046, NORDITA 2021-028, QMUL-PH-21-17",
    doi = "10.1007/JHEP07(2021)169",
    journal = "JHEP",
    volume = "07",
    pages = "169",
    year = "2021"
}

@article{Bern:2021dqo,
    author = "Bern, Zvi and Parra-Martinez, Julio and Roiban, Radu and Ruf, Michael S. and Shen, Chia-Hsien and Solon, Mikhail P. and Zeng, Mao",
    title = "{Scattering Amplitudes and Conservative Binary Dynamics at ${\cal O}(G^4)$}",
    eprint = "2101.07254",
    archivePrefix = "arXiv",
    primaryClass = "hep-th",
    reportNumber = "CALT-TH-2021-004, FR-PHENO-2021-03, OUTP-21-03P",
    doi = "10.1103/PhysRevLett.126.171601",
    journal = "Phys. Rev. Lett.",
    volume = "126",
    number = "17",
    pages = "171601",
    year = "2021"
}

@article{Driesse:2024xad,
    author = "Driesse, Mathias and Jakobsen, Gustav Uhre and Mogull, Gustav and Plefka, Jan and Sauer, Benjamin and Usovitsch, Johann",
    title = "{Conservative Black Hole Scattering at Fifth Post-Minkowskian and First Self-Force Order}",
    eprint = "2403.07781",
    archivePrefix = "arXiv",
    primaryClass = "hep-th",
    reportNumber = "HU-EP-24/08-RTG, CERN-TH-2024-032",
    doi = "10.1103/PhysRevLett.132.241402",
    journal = "Phys. Rev. Lett.",
    volume = "132",
    number = "24",
    pages = "241402",
    year = "2024"
}

@article{Dlapa:2025biy,
    author = {Dlapa, Christoph and K{\"a}lin, Gregor and Liu, Zhengwen and Porto, Rafael A.},
    title = "{Local-in-Time Conservative Binary Dynamics at Fifth Post-Minkowskian and First Self-Force Orders}",
    eprint = "2506.20665",
    archivePrefix = "arXiv",
    primaryClass = "hep-th",
    journal = "",
    reportNumber = "DESY 25-089",
    month = "6",
    year = "2025"
}

@article{Damour:2025uka,
    author = "Damour, Thibault and Nagar, Alessandro and Placidi, Andrea and Rettegno, Piero",
    title = "{A novel Lagrange-multiplier approach to the effective-one-body dynamics of binary systems in post-Minkowskian gravity}",
    eprint = "2503.05487",
    archivePrefix = "arXiv",
    primaryClass = "gr-qc",
    journal = "",
    month = "3",
    year = "2025"
}

@article{Dlapa:2024cje,
    author = {Dlapa, Christoph and K{\"a}lin, Gregor and Liu, Zhengwen and Porto, Rafael A.},
    title = "{Local in Time Conservative Binary Dynamics at Fourth Post-Minkowskian Order}",
    eprint = "2403.04853",
    archivePrefix = "arXiv",
    primaryClass = "hep-th",
    reportNumber = "DESY 24-029",
    doi = "10.1103/PhysRevLett.132.221401",
    journal = "Phys. Rev. Lett.",
    volume = "132",
    number = "22",
    pages = "221401",
    year = "2024"
}

@article{DiVecchia:2023frv,
    author = "Di Vecchia, Paolo and Heissenberg, Carlo and Russo, Rodolfo and Veneziano, Gabriele",
    title = "{The gravitational eikonal: From particle, string and brane collisions to black-hole encounters}",
    eprint = "2306.16488",
    archivePrefix = "arXiv",
    primaryClass = "hep-th",
    reportNumber = "CERN-TH-2023-108, NORDITA 2023-026, QMUL-PH-23-09, UUITP-14/23",
    doi = "10.1016/j.physrep.2024.06.002",
    journal = "Phys. Rept.",
    volume = "1083",
    pages = "1--169",
    year = "2024"
}

@article{Damour:2019lcq,
    author = "Damour, Thibault",
    title = "{Classical and quantum scattering in post-Minkowskian gravity}",
    eprint = "1912.02139",
    archivePrefix = "arXiv",
    primaryClass = "gr-qc",
    doi = "10.1103/PhysRevD.102.024060",
    journal = "Phys. Rev. D",
    volume = "102",
    number = "2",
    pages = "024060",
    year = "2020"
}

@article{Rettegno:2023ghr,
    author = "Rettegno, Piero and Pratten, Geraint and Thomas, Lucy M. and Schmidt, Patricia and Damour, Thibault",
    title = "{Strong-field scattering of two spinning black holes: Numerical relativity versus post-Minkowskian gravity}",
    eprint = "2307.06999",
    archivePrefix = "arXiv",
    primaryClass = "gr-qc",
    doi = "10.1103/PhysRevD.108.124016",
    journal = "Phys. Rev. D",
    volume = "108",
    number = "12",
    pages = "124016",
    year = "2023"
}

@article{Parnachev:2020zbr,
    author = "Parnachev, Andrei and Sen, Kallol",
    title = "{Notes on AdS-Schwarzschild eikonal phase}",
    eprint = "2011.06920",
    archivePrefix = "arXiv",
    primaryClass = "hep-th",
    doi = "10.1007/JHEP03(2021)289",
    journal = "JHEP",
    volume = "03",
    pages = "289",
    year = "2021"
}

@article{Poisson:1994yf,
    author = "Poisson, Eric and Sasaki, Misao",
    title = "{Gravitational radiation from a particle in circular orbit around a black hole. 5: Black hole absorption and tail corrections}",
    eprint = "gr-qc/9412027",
    archivePrefix = "arXiv",
    doi = "10.1103/PhysRevD.51.5753",
    journal = "Phys. Rev. D",
    volume = "51",
    pages = "5753--5767",
    year = "1995"
}

@article{Riva:2021vnj,
    author = "Riva, Massimiliano Maria and Vernizzi, Filippo",
    title = "{Radiated momentum in the post-Minkowskian worldline approach via reverse unitarity}",
    eprint = "2110.10140",
    archivePrefix = "arXiv",
    primaryClass = "hep-th",
    doi = "10.1007/JHEP11(2021)228",
    journal = "JHEP",
    volume = "11",
    pages = "228",
    year = "2021"
}

@article{Saketh:2021sri,
    author = "Saketh, M. V. S. and Vines, Justin and Steinhoff, Jan and Buonanno, Alessandra",
    title = "{Conservative and radiative dynamics in classical relativistic scattering and bound systems}",
    eprint = "2109.05994",
    archivePrefix = "arXiv",
    primaryClass = "gr-qc",
    doi = "10.1103/PhysRevResearch.4.013127",
    journal = "Phys. Rev. Res.",
    volume = "4",
    number = "1",
    pages = "013127",
    year = "2022"
}

@article{Tanaka:1996lfd,
    author = "Tanaka, Takahiro and Tagoshi, Hideyuki and Sasaki, Misao",
    title = "{Gravitational waves by a particle in circular orbits around a Schwarzschild black hole: 5.5 postNewtonian formula}",
    eprint = "gr-qc/9701050",
    archivePrefix = "arXiv",
    reportNumber = "OU-TAP-44, NAOJ-TH-AP-1996-3",
    doi = "10.1143/PTP.96.1087",
    journal = "Prog. Theor. Phys.",
    volume = "96",
    pages = "1087--1101",
    year = "1996"
}

@article{Damour:2016gwp,
    author = "Damour, Thibault",
    title = "{Gravitational scattering, post-Minkowskian approximation and Effective One-Body theory}",
    eprint = "1609.00354",
    archivePrefix = "arXiv",
    primaryClass = "gr-qc",
    doi = "10.1103/PhysRevD.94.104015",
    journal = "Phys. Rev. D",
    volume = "94",
    number = "10",
    pages = "104015",
    year = "2016"
}

@article{Damour:2017zjx,
    author = "Damour, Thibault",
    title = "{High-energy gravitational scattering and the general relativistic two-body problem}",
    eprint = "1710.10599",
    archivePrefix = "arXiv",
    primaryClass = "gr-qc",
    doi = "10.1103/PhysRevD.97.044038",
    journal = "Phys. Rev. D",
    volume = "97",
    number = "4",
    pages = "044038",
    year = "2018"
}

@inproceedings{Buonanno:2022pgc,
    author = "Buonanno, Alessandra and Khalil, Mohammed and O'Connell, Donal and Roiban, Radu and Solon, Mikhail P. and Zeng, Mao",
    title = "{Snowmass White Paper: Gravitational Waves and Scattering Amplitudes}",
    booktitle = "{Snowmass 2021}",
    eprint = "2204.05194",
    archivePrefix = "arXiv",
    primaryClass = "hep-th",
    month = "4",
    year = "2022"
}

@article{Khalaf:2025jpt,
    author = "Khalaf, Majed and Shen, Chia-Hsien and Telem, Ofri",
    title = "{Bound-unbound universality and the all-order semi-classical wave function in Schwarzschild}",
    eprint = "2503.23317",
    archivePrefix = "arXiv",
    primaryClass = "gr-qc",
    doi = "10.1007/JHEP10(2025)063",
    journal = "JHEP",
    volume = "10",
    pages = "063",
    year = "2025"
}

@article{LIGOScientific:2025rsn,
    author = "Abac, A. G. and others",
    collaboration = "LIGO Scientific, VIRGO, KAGRA",
    title = "{GW231123: A Binary Black Hole Merger with Total Mass 190{\textendash}265 M$_{⊙}$}",
    eprint = "2507.08219",
    archivePrefix = "arXiv",
    primaryClass = "astro-ph.HE",
    reportNumber = "DCC: P2500026-v6, DCC: P2500026-v8",
    doi = "10.3847/2041-8213/ae0c9c",
    journal = "Astrophys. J. Lett.",
    volume = "993",
    number = "1",
    pages = "L25",
    year = "2025"
}

@article{Barack:2018yvs,
    author = "Barack, Leor and Pound, Adam",
    title = "{Self-force and radiation reaction in general relativity}",
    eprint = "1805.10385",
    archivePrefix = "arXiv",
    primaryClass = "gr-qc",
    doi = "10.1088/1361-6633/aae552",
    journal = "Rept. Prog. Phys.",
    volume = "82",
    number = "1",
    pages = "016904",
    year = "2019"
}

@article{Pretorius:2007jn,
    author = "Pretorius, Frans and Khurana, Deepak",
    editor = "Campanelli, Manuela and Rezzolla, Luciano",
    title = "{Black hole mergers and unstable circular orbits}",
    eprint = "gr-qc/0702084",
    archivePrefix = "arXiv",
    doi = "10.1088/0264-9381/24/12/S07",
    journal = "Class. Quant. Grav.",
    volume = "24",
    pages = "S83--S108",
    year = "2007"
}

@article{Quinn:1999kj,
    author = "Quinn, Theodore C. and Wald, Robert M.",
    title = "{Energy conservation for point particles undergoing radiation reaction}",
    eprint = "gr-qc/9903014",
    archivePrefix = "arXiv",
    doi = "10.1103/PhysRevD.60.064009",
    journal = "Phys. Rev. D",
    volume = "60",
    pages = "064009",
    year = "1999"
}

@article{Warburton:2025ymy,
    author = "Warburton, Niels",
    title = "{Gravitational radiation from hyperbolic orbits: comparison between self-force, post-Minkowskian, post-Newtonian, and numerical relativity results}",
    eprint = "2512.02274",
    archivePrefix = "arXiv",
    primaryClass = "gr-qc",
    journal = "",
    month = "12",
    year = "2025"
}

@article{Galtsov:1982hwm,
    author = "Gal'tsov, D. V.",
    title = "{Radiation reaction in the Kerr gravitational field}",
    doi = "10.1088/0305-4470/15/12/025",
    journal = "J. Phys. A",
    volume = "15",
    pages = "3737--3749",
    year = "1982"
}

@article{LIGOScientific:2025slb,
    author = "Abac, A. G. and others",
    collaboration = "LIGO Scientific, VIRGO, KAGRA",
    title = "{GWTC-4.0: Updating the Gravitational-Wave Transient Catalog with Observations from the First Part of the Fourth LIGO-Virgo-KAGRA Observing Run}",
    eprint = "2508.18082",
    archivePrefix = "arXiv",
    primaryClass = "gr-qc",
    journal = "",
    reportNumber = "LIGO-P2400386",
    month = "8",
    year = "2025"
}

@article{LISA:2024hlh,
    author = "Colpi, Monica and others",
    collaboration = "LISA",
    title = "{LISA Definition Study Report}",
    eprint = "2402.07571",
    archivePrefix = "arXiv",
    journal = "",
    primaryClass = "astro-ph.CO",
    month = "2",
    year = "2024"
}

@misc{ZenodoMma,
  author       = {Barack, Leor and
                  Gonzo, Riccardo and
                  Leather, Benjamin and
                  Long, Oliver and
                  Warburton, Niels},
  title        = {{ancillary.m}},
  year         = {2026},
  howpublished = {Zenodo},
  doi          = {10.5281/zenodo.18553926},
  url          = {https://doi.org/10.5281/zenodo.18553926}
}

@misc{ZenodoCirc,
  author       = {Barack, Leor and
                  Gonzo, Riccardo and
                  Leather, Benjamin and
                  Long, Oliver and
                  Warburton, Niels},
  title        = {{UnstableCircularOrbitFluxes.dat}},
  year         = {2026},
  howpublished = {Zenodo},
  doi          = {10.5281/zenodo.18553926},
  url          = {https://doi.org/10.5281/zenodo.18553926}
}

@misc{ZenodoAbs,
  author       = {Barack, Leor and
                  Gonzo, Riccardo and
                  Leather, Benjamin and
                  Long, Oliver and
                  Warburton, Niels},
  title        = {{AbsorbedEnergyScattering.dat}},
  year         = {2026},
  howpublished = {Zenodo},
  doi          = {10.5281/zenodo.18553926},
  url          = {https://doi.org/10.5281/zenodo.18553926}
}

@misc{ZenodoRad,
  author       = {Barack, Leor and
                  Gonzo, Riccardo and
                  Leather, Benjamin and
                  Long, Oliver and
                  Warburton, Niels},
  title        = {{RadiatedEnergyScattering.dat}},
  year         = {2026},
  howpublished = {Zenodo},
  doi          = {10.5281/zenodo.18553926},
  url          = {https://doi.org/10.5281/zenodo.18553926}
}

@article{OShaughnessy:2002tbu,
    author = "O'Shaughnessy, Richard W.",
    title = "{Transition from inspiral to plunge for eccentric equatorial Kerr orbits}",
    eprint = "gr-qc/0211023",
    archivePrefix = "arXiv",
    doi = "10.1103/PhysRevD.67.044004",
    journal = "Phys. Rev. D",
    volume = "67",
    pages = "044004",
    year = "2003"
}

@article{Albanesi:2024xus,
    author = "Albanesi, Simone and Rashti, Alireza and Zappa, Francesco and Gamba, Rossella and Cook, William and Daszuta, Boris and Bernuzzi, Sebastiano and Nagar, Alessandro and Radice, David",
    title = "{Scattering and dynamical capture of two black holes: Synergies between numerical and analytical methods}",
    eprint = "2405.20398",
    archivePrefix = "arXiv",
    primaryClass = "gr-qc",
    doi = "10.1103/PhysRevD.111.024069",
    journal = "Phys. Rev. D",
    volume = "111",
    number = "2",
    pages = "024069",
    year = "2025"
}

\end{document}